\documentclass[10pt,letterpaper,twocolumn,english,final,floatfix,showpacs,showkeys,citesort,preprintnumbers,prd,nofootinbib,superscriptaddress]{revtex4-1}
\usepackage[T1]{fontenc}
\usepackage[latin1]{inputenc}
\usepackage{color}
\usepackage{babel}
\usepackage{float}
\usepackage{calc}
\usepackage{amsmath}
\usepackage{amssymb}
\usepackage{graphicx}
\usepackage{esint}
\usepackage[unicode=true,pdfusetitle,
 bookmarks=true,bookmarksnumbered=false,bookmarksopen=false,
 breaklinks=false,pdfborder={0 0 1},colorlinks=false]
 {hyperref}

\usepackage{paralist}

\makeatletter

\pdfpageheight\paperheight
\pdfpagewidth\paperwidth

 \@ifundefined{textcolor}{}
 {%
   \definecolor{BLACK}{gray}{0}
   \definecolor{WHITE}{gray}{1}
   \definecolor{RED}{rgb}{1,0,0}
   \definecolor{GREEN}{rgb}{0,1,0}
   \definecolor{BLUE}{rgb}{0,0,1}
   \definecolor{CYAN}{cmyk}{1,0,0,0}
   \definecolor{MAGENTA}{cmyk}{0,1,0,0}
   \definecolor{YELLOW}{cmyk}{0,0,1,0}
 }

\@ifundefined{definecolor}
 {\usepackage{color}}{}
\usepackage{babel}

\newcommand{\be}{\begin{equation}}
\newcommand{\ee}{\end{equation}}
\newcommand{\ba}{\begin{eqnarray}}
\newcommand{\ea}{\end{eqnarray}}

\newcommand{\der}{\ensuremath{{\operatorname{d}}}}

\@ifundefined{showcaptionsetup}{}{%
 \PassOptionsToPackage{caption=false}{subfig}}
\usepackage{subfig}
\makeatother

\begin{document} 

\preprint{
\vbox{
\null \vspace{0.3in}
\hbox{DO-TH 14/28}
\hbox{LPSC-14-120}
\hbox{SMU-HEP-14-05}
}
}

\title{\null \vspace{0.5in}
Comparison of the ANP model with the data for neutrino induced single pion production from 
the MiniBooNE and MINER$\nu$A experiments}

\author{J.~Y.~Yu}
\thanks{yu@physics.smu.edu}
\affiliation{Southern Methodist University,Dallas, TX 75275, USA}

\author{E.~A.~Paschos}
\thanks{paschos@physik.uni-dortmund.de}
\affiliation{Department of Physics, Technical University of Dortmund, D-44221 Dortmund, Germany}

\author{I.~Schienbein}
\thanks{ingo.schienbein@lpsc.in2p3.fr}
\affiliation{Laboratoire de Physique Subatomique et de Cosmologie, Universit\'e Grenoble-Alpes, CNRS/IN2P3,
                   53 avenue des Martyrs, F-38026 Grenoble, France}

\keywords{QCD, Neutrino interactions, Single pion production}

\pacs{change 12.38.-t,13.15.+g,13.60.-r,24.85.+p}

\begin{abstract}
We present theoretical predictions in the framework of the ANP model
for single pion production ($\pi^+, \pi^0$)
in $\nu_\mu$ and $\bar\nu_\mu$ scattering off mineral oil and plastic.
Our results for the total cross sections and flux averaged differential distributions
are compared to all available data of the MiniBooNE and MINER$\nu$A experiments.
While our predictions slightly undershoot the MiniBooNE data they 
reproduce the normalization of the MINER$\nu$A data for the kinetic energy
distribution.
For the dependence on the polar angle we reproduce the shape of the
arbitrarily normalized data.
\end{abstract}

\maketitle
\tableofcontents{}

\newpage{}


\section{Introduction}
A good understanding of neutrino--nucleus interactions in the few GeV and sub-GeV energy range 
is a key ingredient  for precision measurements of the properties of neutrino oscillations which will allow
to discover or constrain the existence of sterile neutrinos, CP-violation in the leptonic sector or to improve
the uncertainty on the mixing angle $\theta_{13}$.
Several models to calculate these cross sections in the low and intermediate energy regions
have been discussed in the literature and a summary of recent developments can be found in \cite{Morfin:2012kn}.

Recently, the MiniBooNE collaboration has published 
single pion production cross section data at $E_\nu <2$ GeV for 
charged current (CC) $\pi^+$ production \cite{AguilarArevalo:2010bm},  CC $\pi^0$ production \cite{AguilarArevalo:2010xt},
and (anti-)neutrino induced neutral current (NC) $\pi^0$ production \cite{AguilarArevalo:2009ww}.  
Since these data are measured in mineral oil  (CH$_2$), 
it is not meaningful to compare them to free nucleon cross section models  
due to nuclear effects such as in-medium modifications of the free nucleon cross sections
and final state interaction (FSI) effects.

The MiniBooNE results have triggered a lot of interest in the community
and some of the existing theoretical models for single pion production \cite{Lalakulich:2012cj,Hernandez:2013jka} 
which include FSI and in-medium modifications have been compared with 
the MiniBooNE data and have the tendency to undershoot them.

In another recent publication \cite{Paschos:2012tr}, the differential cross section 
$\der \sigma/\der Q^2$ has been calculated in the small $Q^2$ region
including FSI effects using the ANP model and has been compared with MiniBooNE data for CC $\pi^+$ and $\pi^0$ production.
A good agreement with the data was found demonstrating the validity of the ANP model without the need of any modifications.

The work in  \cite{Paschos:2012tr} is limited to the small $Q^2$ region.
In this paper, we compute the total and differential cross sections for CC and NC single pion production 
including  nuclear effects (Pauli suppression, Fermi motion, charge exchange, absorption)
in the framework of our earlier publications \cite{Paschos:2000be,Schienbein:2003sm,Paschos:2003qr,Paschos:2004qh,Paschos:2004md} 
and perform a comprehensive comparison with all of the above mentioned MiniBooNE data. 
In addition, we perform a comparison with the most recent results on CC1$\pi^+$ production in plastic (CH)
from the MINER$\nu$A experiment  \cite{Eberly:2014mra}.
This work will serve as a reference for a future study where we recalculate the cross sections including additional
resonant and non-resonant contributions \cite{psy-winp}.

The rest of this paper is organized as follows. 
In Sec.\ \ref{sec:theory} we briefly review our model for neutrino induced pion production for the cases of free nucleon and nuclear targets.
In Secs.\ \ref{sec:miniboone} and \ref{sec:minerva}  we present our numerical results for CC and NC single pion production
and perform a detailed comparison with the complete set of available MiniBooNE and MINER$\nu$A data, respectively.
Finally, in Sec.\ \ref{sec:summary} we summarize the main results and present our conclusions.  

\section{Neutrino induced single pion production}
\label{sec:theory}

The theory for the production of the $P_{33}$ (or $\Delta(1232)$)  resonance  is well understood 
for the free nucleon case and  several calculations are available in the literature 
\cite{Fogli:1979cz,Fogli:1979qj,Schreiner:1973mj,Rein:1980wg,Lalakulich:2012cj,Hernandez:2013jka} 
showing agreement with the experimental results from the Argonne National Laboratory (ANL) \cite{Radecky:1981fn} and the 
Brookhaven National Laboratory (BNL) \cite{Kitagaki:1990vs}.
For our purpose, we employ the formalism of Schreiner and von Hippel (SvH) \cite{Schreiner:1973mj}.
Furthermore, for the higher resonances $N(1440)P_{11}$ and $N(1535)S_{11}$ we follow the article \cite{Paschos:2000be}.
For single pion production in neutrino--nucleus scattering we then use the ANP model.
Since the details of these calculations have been discussed already in our earlier publications
\cite{Paschos:2000be,Schienbein:2003sm,Paschos:2003qr,Paschos:2004qh,Paschos:2004md} 
we will only briefly summarize our formalism in the following.
  
\subsection{Free nucleon case}
The triple-differential cross section for $\Delta$ resonance production 
can be obtained from the  fully differential cross section ${\frac{\der^4\sigma}{\der Q^2\der W \der \Omega_\pi}}$
given in  \cite{Schreiner:1973mj} by integrating over the azimuthal angle $\phi_\pi$ and performing a change of variables:
\begin{eqnarray} 
{\frac{\der^3\sigma}{\der Q^2\der W \der E_\pi}} &=&
\frac{1}{\beta\gamma|p_\pi^{CMS}|}\frac{WG_F^2}{16\pi M_N^2} \times
\\
&&
\Sigma_{i=1}^3 \left[K_i\tilde{W}_i-\frac{1}{2}K_i D_i(3 \cos^2\theta_\pi-1)\right]\, .
\nonumber
\end{eqnarray}
Here, $Q^2$ is the virtuality of the exchange-boson, $W$ the invariant mass of the final state pion-nucleon system, and
$E_\pi$ the energy of the pion in the laboratory frame.
Furthermore, $G_F$ is the Fermi constant, $M_N$ the nucleon mass, $\beta$ the velocity of the $\Delta$ resonance in the laboratory frame, 
$\gamma$ the corresponding Lorentz factor, and $p_\pi^{CMS}$ the modulus of the pion three-momentum in the center-of-mass frame.
The $K_i$ are kinematic factors and the dynamics of the process is contained in the structure functions $\tilde{W}_i$ and $D_i$.
For the calculation of the two higher resonances $P_{11}$ and $S_{11}$ we take the triple-differential cross section 
from \cite{Paschos:2000be}.
The total cross section is then obtained by integrating over the valid kinematical ranges of $Q^2$, the pion energy $E_\pi$, and the 
invariant mass in the interval $W \in [M_N+m_\pi,W_{\rm max}]$ where $W_{\rm max}$ will be specified in Secs. \ref{sec:miniboone} and
\ref{sec:minerva} for the different experiments.

\begin{figure}[t]
\centering
\includegraphics[width=0.45\textwidth]{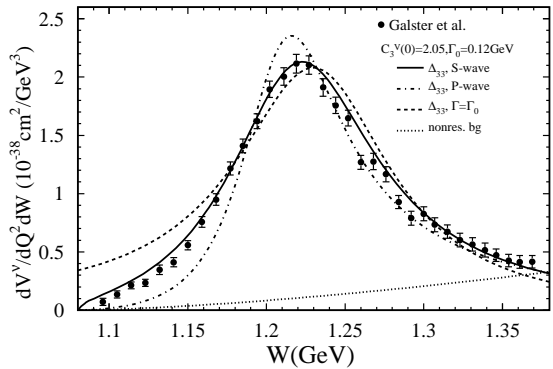} 
\vspace*{-1.0cm}
\caption{\sf 
Cross section $d^2 V^\nu/dQ^2dW$ for electroproduction in the $\Delta$ resonance region
in comparison with data from Galster et al.\ \cite{Galster:1972rh}. 
The solid, dot-dashed and dashed lines have been obtained with a running s-wave width, a running p-wave width,
and a constant width, respectively. Also shown is the non-resonant background (dotted line).
The resonant curves have been obtained using $C_3^V(0)=2.05$ and $\Gamma_0=0.120$ GeV.}
 \label{fig:sigw1}
\end{figure}

The $\Delta$ resonance is modeled by a Breit-Wigner distribution with a running width
\begin{equation}
\Gamma(W) = \Gamma_0 \left[\frac{q_\pi(W)}{q_\pi(M_\Delta)}\right]^J
\end{equation}
with $J=1$ (s-wave) and $\Gamma_0=0.120$ GeV  \cite{Paschos:2003qr}.
As is visible in Fig.\ \ref{fig:sigw1}, the s-wave curve gives a very good description
of the data by Galster et al.~\cite{Galster:1972rh} for the hadronic invariant mass distribution 
$W$ in electron-proton scattering. 
A similar figure has been shown in \cite{Paschos:2003qr} and we refer to this article for the details.
However, it is instructive to see the corresponding curves for a p-wave form ($J=3$) and for a constant width ($J=0$)
using the same input parameters ($\Gamma_0, C_3^V(0)$). As can be seen they give an inferior description of the data.
Furthermore, we use modified dipole form factors which have been introduced in reference \cite{Paschos:2003qr}. 
The form factors for the higher  $P_{11}$ and $S_{11}$ resonances  can also be found in \cite{Paschos:2003qr}. 
Apart from the mass and the width of the $\Delta$ resonance, this model employs four free parameters to
describe the production of the $\Delta$ resonance:
The values of the vector form factors $C_3^V(0)$ at $Q^2=0$, the vector mass $M_V$,
the value of the axial vector form factor $C_5^A(0)$ at $Q^2=0$, and the axial vector mass $M_A$.
The latter two parameters
were fitted  to the flux averaged $Q^2$-differential cross section measured at BNL \cite{Kitagaki:1990vs} in the region 
$Q^2>0.15$ GeV$^2$.
Plots of the form factors are very close to subsequent determinations using electroproduction data and PCAC
\cite{Lalakulich:2006sw}.
For convenience, we summarize the input parameters 
concerning the $\Delta$ resonance production in Table \ref{tab:input}.
We note that $C_5(Q^2)$ was also fitted with $C_5^A(0)$ and $M_A$ as free parameters in a dipole and also a modified dipole parametrization 
of the form factors in \cite{Graczyk:2009qm}; they found values close to the ones in Table \ref{tab:input}.
In Secs.\ \ref{sec:miniboone} and \ref{sec:minerva}, we use these parameters to compute
our cross sections for the MiniBooNE and the MINER$\nu$A data.
It should be noted that these results constitute real predictions
using the original framework as outlined in \cite{Paschos:2003qr}.
The only minor difference is that we do {\em not} neglect the muon mass and 
we therefore include the contribution from the form factor $C_6^A$ for
which we use
\begin{equation}
C_6^A(Q^2) = C_5^A(Q^2) \frac{M_N^2}{Q^2+m_\pi^2}\, .
\end{equation}
Having summarized our framework for single pion production in the case of a free nucleon
target we now turn to a discussion of the nuclear case.

\begin{table}[!t]
\center
\begin{tabular}{|c|c|c|c|c|c|}
\hline
$m_\Delta$ [GeV] & $\Gamma_0$ [GeV] & $C_3^V(0)$ & $M_V [GeV]$ & $C_5^A(0)$ & $M_A$ [GeV] \\\hline
1.232 & 0.120 & 1.95 & 0.84 & 1.2 & 1.05\\
\hline
\end{tabular}
\caption{Input parameters for the $\Delta$ resonance production \protect\cite{Paschos:2003qr}.}
\label{tab:input}
\end{table}

\subsection{ANP model}
In the 70's S.\ L.\ Adler, S.\ Nussinov and E.\ A.\ Paschos proposed the ANP model 
to describe the nuclear corrections to leptonic pion production in the $\Delta$ resonance region in a simple way \cite{Adler:1974qu}. 
A more recent account of the ANP model can also be found in \cite{Paschos:2000be,Schienbein:2003sm,Paschos:2004qh}.
Here we summarize its essential features to make the paper self-contained.

A key ingredient of this model is the assumption
that the process can be factorized into two independent steps.
In step 1, the neutrino interacts with one of the nucleons inside the nuclear target producing a pion.
This cross section is reduced by the Pauli blocking factor and broadened by the Fermi motion.
In step 2, the subsequent rescattering of the pions is described by a transport matrix
by an absorption term and the pion--proton and pion--neutron cross sections including charge exchange effects.
The distributions of protons and neutrons in the nucleus are proportional to the nucleon density.
The mathematical transport problem was solved exactly, as well as in approximate geometrical cases, 
where the scattered pions were all projected in the forward- and backward-hemisphere.
Comparison of the exact and approximate multiple scattering solutions shows that they are very close to each other
(see Eqs.\ (B4)--(B6) and Table IX in Ref.\ \cite{Adler:1974qu}).
This means that the pions after the multiple scattering essentially preserve the direction of the first step.
The initial pion yields within the nucleus $\Sigma^p_{\pi^+,i} = d \sigma(\nu p \to \mu^- p \pi^+)$,
$\Sigma^n_{\pi^+,i} = d \sigma(\nu n \to \mu^- n \pi^+)$ and
$\Sigma^n_{\pi^0,i} = d \sigma(\nu n \to \mu^- p \pi^0)$ will produce the final yields
(denoted by the subscript 'f'):
\begin{equation}
\left(\begin{array}{c}
\Sigma_{\pi^+}
\\
\Sigma_{\pi^0}
\\
\Sigma_{\pi^-}
\end{array}\right)_{\rm f}
= M(_6C^{12})\ 
\left(\begin{array}{c}
6 \Sigma^p_{\pi^+} + 6 \Sigma^n_{\pi^+}
\\
6 \Sigma^n_{\pi^0}
\\
0
\end{array}\right)_{\rm i}
\quad .
\label{eq:fac}
\end{equation}

It should be noted that this approach is quite general, only relying on the factorization assumption, and the elements of the transport matrix can in principle 
be extracted from experiment.
On the other hand, in the ANP model, the elements of this matrix are calculated providing predictions which can be tested experimentally.
Four our numerical analysis we use the following matrices 
for 15\%, 20\%, and 25\% effective absorption (see \cite{Schienbein:2003sm,Paschos:2007pe}):
\\

\noindent\underline{15\% absorption}
\begin{equation}
M(_{6}C^{12}) = \overline{A}(Q^2) \left(\begin{array}{ccc}
{0.817} & {0.141} & {0.041} \\
{0.141} & {0.718} & {0.141} \\
{0.041} & {0.141} & {0.817}
\end{array}\right)
\label{eq:MC15}
\end{equation}
with $\overline{A}(Q^2)  = g(Q^2,W=1.2\ {\rm GeV}) \times 0.847$ .\\

\noindent\underline{20\% absorption}
\begin{equation}
M(_{6}C^{12}) = \overline{A}(Q^2) \left(\begin{array}{ccc}
{0.829}& {0.134} & {0.037} \\
{0.134}& {0.731} & {0.134} \\
{0.037}& {0.134} & {0.829}
\end{array}\right)
\label{eq:MC20}
\end{equation}
with $\overline{A}(Q^2)  = g(Q^2,W=1.2\ {\rm GeV}) \times 0.809$ .\\

\noindent\underline{25\% absorption}
\begin{equation}
M(_{6}C^{12}) = \overline{A}(Q^2) \left(\begin{array}{ccc}
{0.840}& {0.127} & {0.032} \\
{0.127}& {0.745} & {0.127} \\
{0.032}& {0.127} & {0.840}
\end{array}\right)
\label{eq:MC25}
\end{equation}
with $\overline{A}(Q^2)  = g(Q^2,W=1.2\ {\rm GeV}) \times 0.752$ .\\

These matrices have been obtained by averaging over $W$ with the
leading $W$-dependence coming from the $\Delta$-resonance contribution;
for details see \cite{Paschos:2007pe} and Eq.\ (47) in Ref.\ \cite{Adler:1974qu}.
However, we did not include the Pauli blocking factor $g(Q^2,W)$ in the averaging procedure since its
dependence on $W$ is very weak and we evaluate it instead at a fixed $W=1.2$ GeV
using the expression in \cite{Paschos:2003qr}.
Note also, that the matrices in Eqs.\ \eqref{eq:MC15}--\eqref{eq:MC25} resemble the ones
in Eqs.\ (B1)--(B3) in Ref.\ \cite{Paschos:2007pe} which have been obtained after averaging also
over $Q^2$ so that $A(Q^2)$ is replaced by a constant number $\overline{\overline{A_p}}$.

We apply this formalism to the MiniBooNE and the MINER$\nu$A data. The target in the experiments
is the molecule CH$_2$ (MiniBooNE) and CH (MINER$\nu$A), respectively. The final pion yields are then 
obtained by adding the yields in carbon to the corresponding free proton yields:
\begin{equation}
\left(\begin{array}{c}
\Sigma_{\pi^+}
\\
\Sigma_{\pi^0}
\\
\Sigma_{\pi^-}
\end{array}\right)_{\rm f}
= M(_6C^{12})\ 
\left(\begin{array}{c}
6 \Sigma^p_{\pi^+} + 6 \Sigma^n_{\pi^+}
\\
6 \Sigma^n_{\pi^0}
\\
0
\end{array}\right)_{\rm i}
+ k \left(\begin{array}{c}
\Sigma^p_{\pi^+} 
\\
0
\\
0
\end{array}\right) ,
\label{eq:final}
\end{equation}
where $k=2$ ($k=1$) for MiniBooNE (MINER$\nu$A).

While in our study we only consider transitions due to the multiple scattering from a single pion $\pi^{0,\pm}$ to a single pion $\pi^{0,\pm}$,
so that the charge-exchange matrix is a $3\times 3$ matrix, it is possible to include more channels, see \cite{Bolognese:1978yz}
and \cite{Paschos:2012tr}.

We note that other groups include additional in-medium modifications of single pion production caused by a change of the mass and the width
of the $\Delta$ resonance in the nuclear environment \cite{Lalakulich:2012cj,Hernandez:2013jka}.
In the ANP model, concerning step 2, these effects are included in the effective charge exchange and absorption cross sections
entering the calculation of the ANP matrix. In our study,  we use the matrices
in Eqs.\ \eqref{eq:MC15}--\eqref{eq:MC25}
for different amounts of absorption to gauge the associated theoretical uncertainty.
In addition, medium modified parameters could in principle affect the cross section in step 1.
We will discuss this in our concluding remarks.

\section{Comparison with MiniBooNE data}
\label{sec:miniboone}

\begin{figure*}[thp!]
\centering
\includegraphics[width=0.45\textwidth]{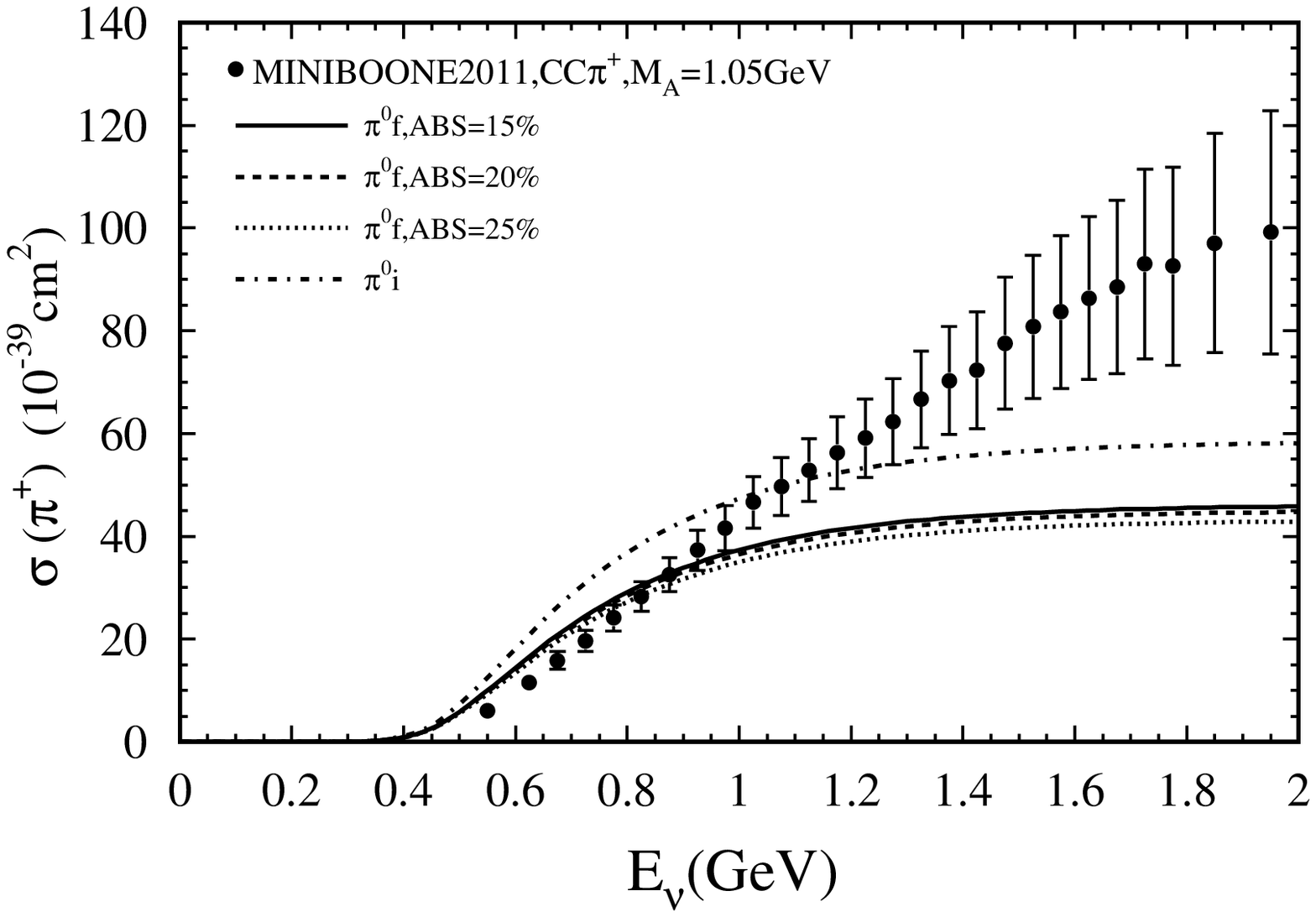} 
\includegraphics[width=0.45\textwidth]{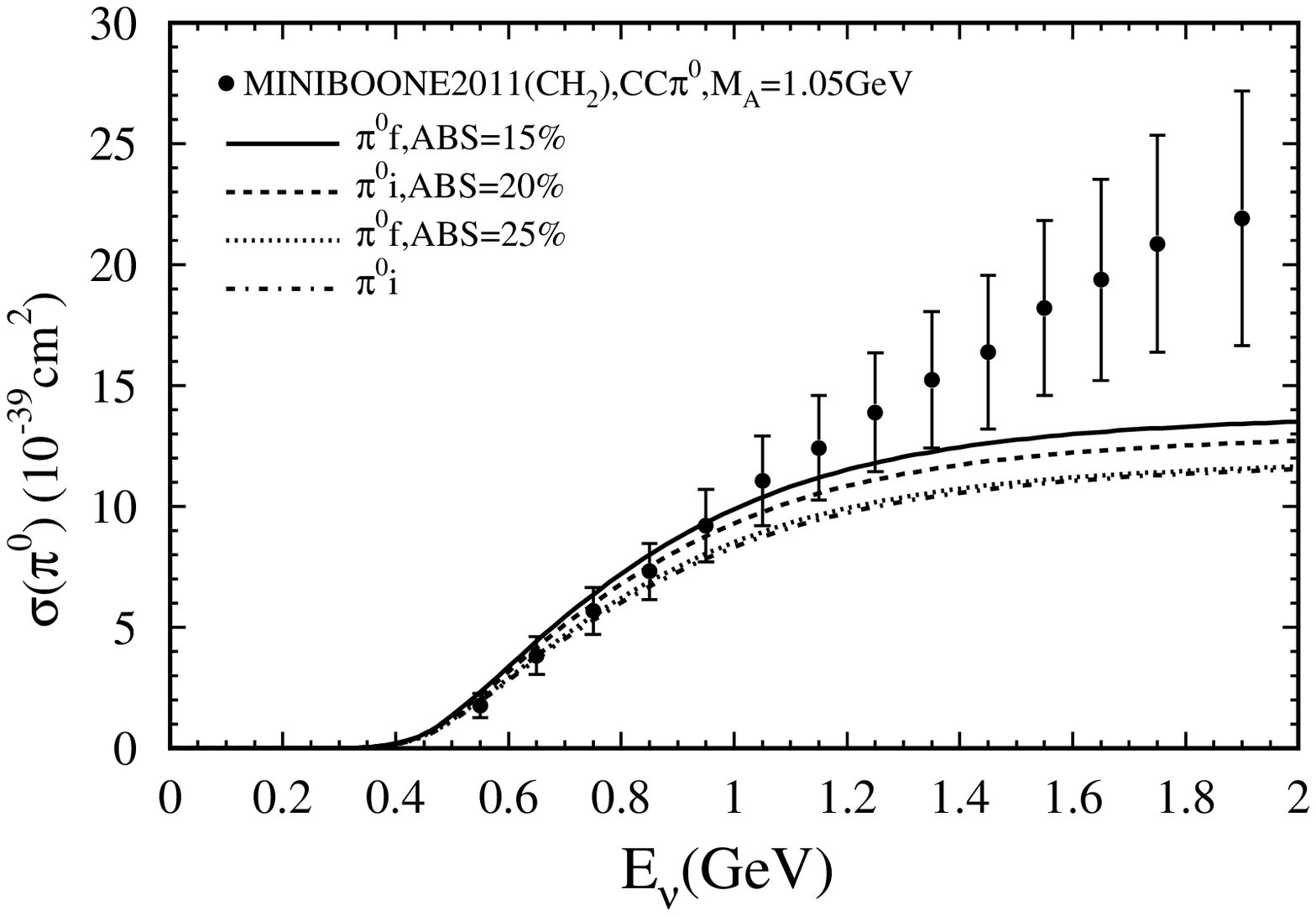} 
\vspace*{-1.0cm}
\caption{\sf 
Total cross sections for CC$1\pi^+$ (left) and CC$1\pi^0$ (right) production in mineral oil (CH$_2$)
in dependence of the neutrino energy $E_\nu$.
The CC$1\pi^+$ data are from Tab.\ V (Fig.\ 20) in \protect\cite{AguilarArevalo:2010bm}
and the CC$1\pi^0$ data from Tab.\ VI (Fig.\ 8) in \protect\cite{AguilarArevalo:2010xt}.
}
 \label{fig:sigtot}
\end{figure*}

\begin{figure*}[thp!]
\centering
\includegraphics[width=0.45\textwidth]{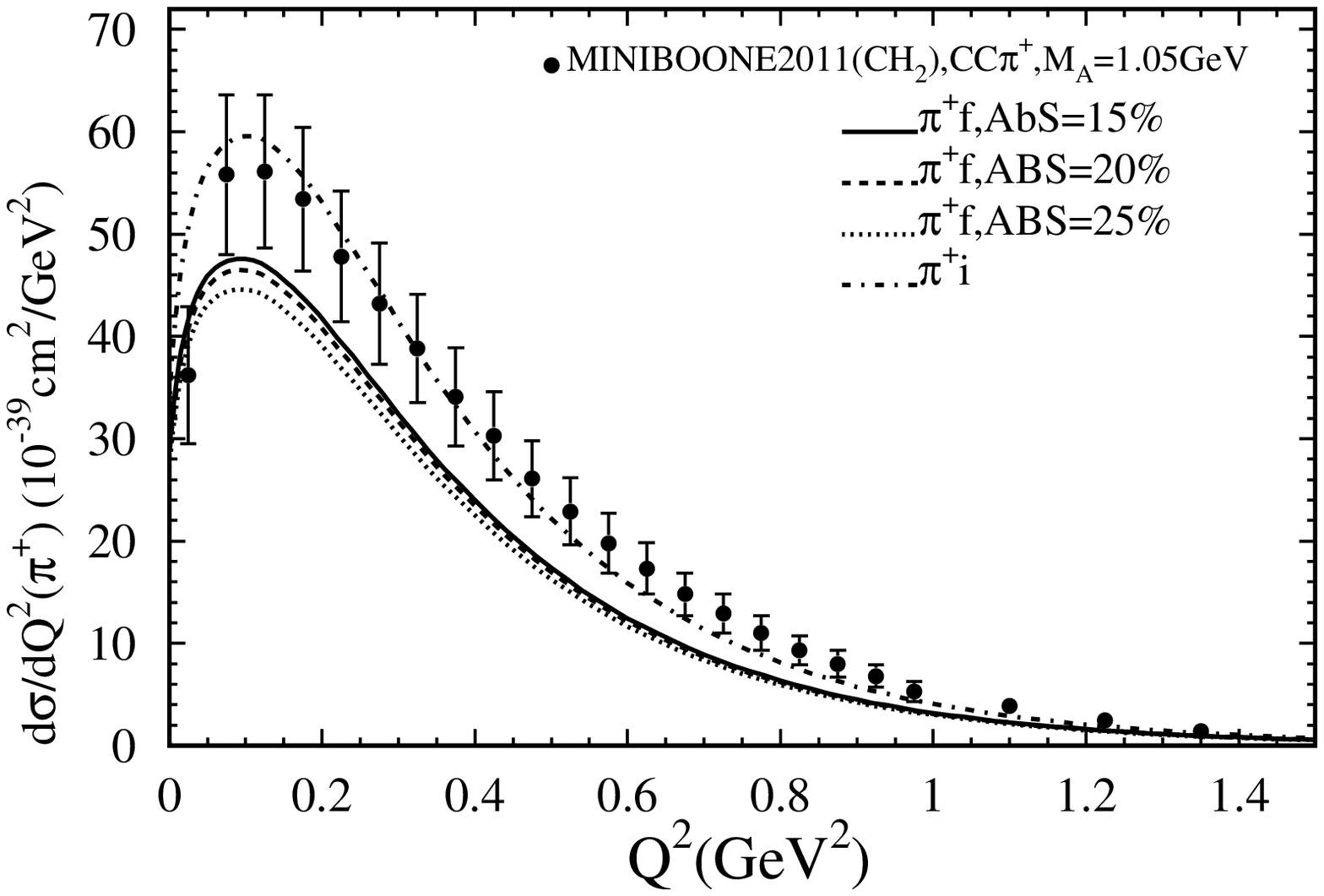} 
\includegraphics[width=0.45\textwidth]{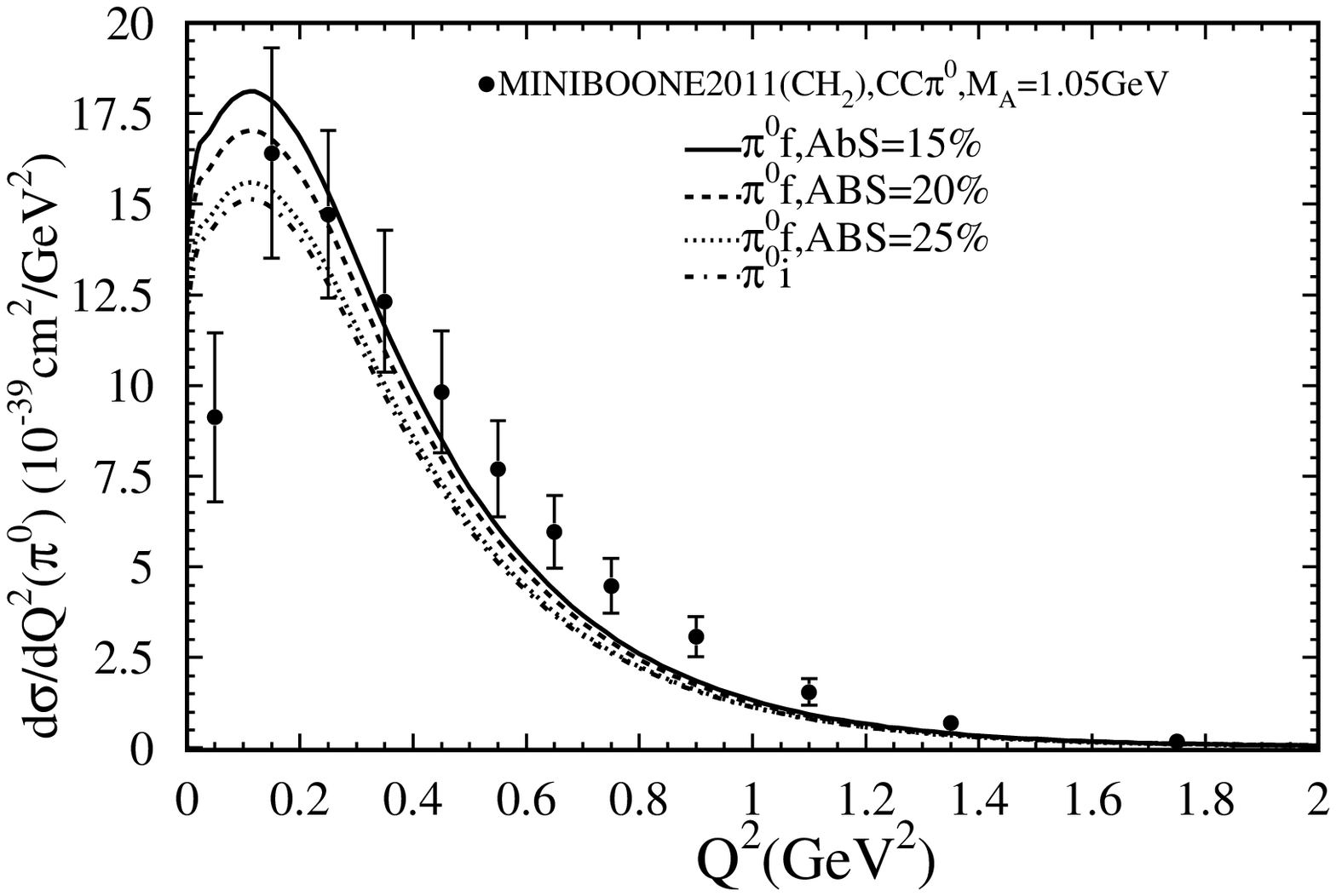} 
\vspace*{-1.cm}
\caption{\sf $Q^2$-differential cross sections for CC$1\pi^+$ (left) 
and CC$1\pi^0$ (right) 
production in mineral oil (CH$_2$) in dependence of $Q^2$.
The CC$1\pi^+$ data are from Tab.\ VII (Fig.\ 21) in \protect\cite{AguilarArevalo:2010bm}
and the CC$1\pi^0$ data from Tab.\ VII (Fig.\ 9) in \protect\cite{AguilarArevalo:2010xt}.
}
\label{fig:sigq2} 
\end{figure*}

\begin{figure*}[thp!]
\centering
\includegraphics[width=0.45\textwidth]{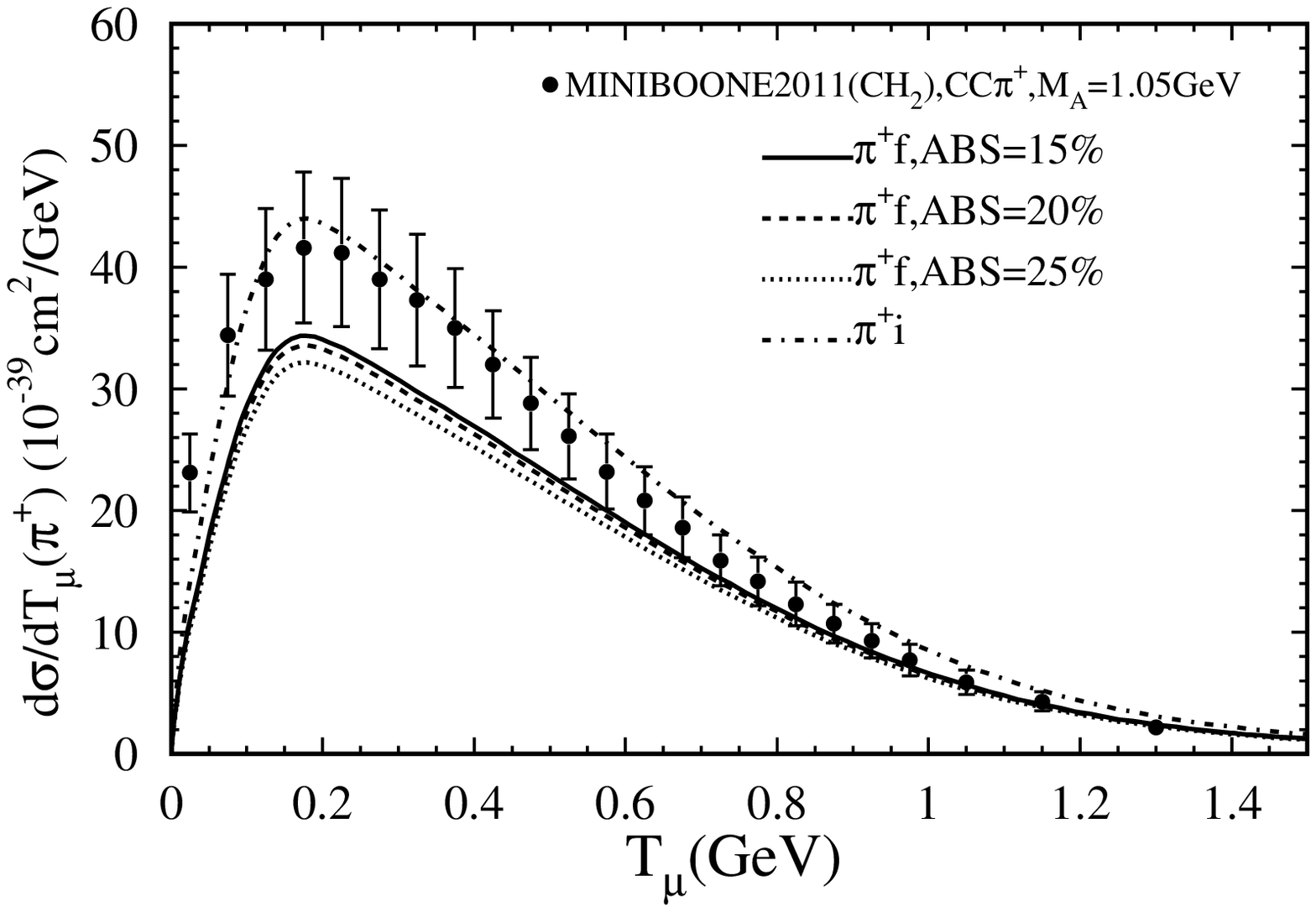}
\includegraphics[width=0.45\textwidth]{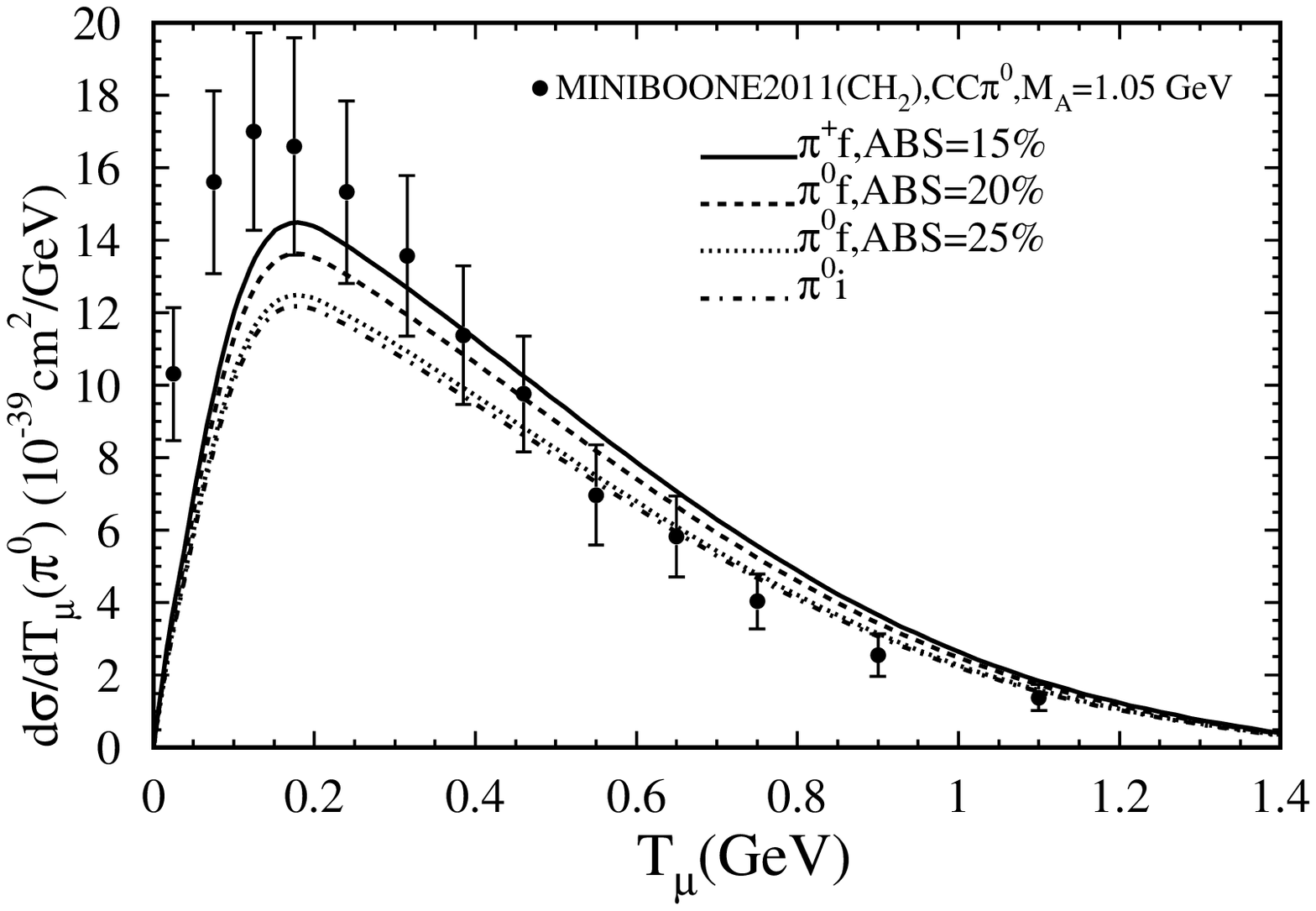}
\vspace*{-1.0cm}
\caption{\sf Same as in Fig.\ \protect\ref{fig:sigq2} for the differential cross sections in dependence of the
kinetic energy of the muon $T_\mu$.
The CC$1\pi^+$ data are from Tab.\ VIII (Fig.\ 22) in \protect\cite{AguilarArevalo:2010bm}
and the CC$1\pi^0$ data from Tab.\ VIII (Fig.\ 10) in \protect\cite{AguilarArevalo:2010xt}.
}
\label{fig:sigtmu}
\end{figure*}

In this section we present our predictions for single pion production in 
$\nu_\mu$ and $\bar \nu_\mu$ scattering off a mineral oil (CH$_2$) target.
We compare our results with recent MiniBooNE data for CC charged pion production (CC$1\pi^+$) \cite{AguilarArevalo:2010bm},
CC neutral pion production (CC$1\pi^0$) \cite{AguilarArevalo:2010xt}, and NC neutral pion production (NC$1\pi^0$) \cite{AguilarArevalo:2009ww}.
For this comparison, we do not include contributions from a non-resonant background, coherent scattering, deep inelastic scattering (DIS)
and the $D_{13}$ resonance which was found to be negligible \cite{Hernandez:2013jka}.\footnote{Note, however, that the authors of 
Ref.\ \protect\cite{Leitner:2008ue} claim that the $D_{13}$ resonance contribution is not negligible at medium neutrino energy.} 
On the other hand, we take into account the small contributions from the $P_{11} $ and $S_{11}$ resonances.

In addition to the total cross section in dependence of the neutrino energy, we calculate flux-averaged differential cross sections.
For this purpose, we use the MiniBooNE flux given in \cite{AguilarArevalo:2009ww,AguilarArevalo:2008yp}
for the CC1$\pi^+$ and NC1$\pi^0$ events covering neutrino energies in the range $E_\nu \in [0,2]$ GeV
and the one in \cite{AguilarArevalo:2010xt} for the CC$1\pi^0$ differential cross sections for neutrino energies
$E_\nu \in [0.5,2]$ GeV.
Following the experimental analysis in Ref.\ \cite{AguilarArevalo:2010bm} we impose a cut on the
invariant mass of the hadronic system $W<1.35$ GeV for the CC1$\pi^+$ events.
For the CC1$\pi^0$ and NC1$\pi^0$ production we use $W_{\rm max}=1.6$ GeV.

In all cases, we present results for the neutrino--nucleon cross section of step 1 denoted $\pi^+_i$ ($\pi^0_i$) in the case of charged 
(neutral) pion production. 
The final results taking into account the final state interactions in step 2 are denoted $\pi^+_f$ respectively $\pi^0_f$.
Each time we present three curves obtained with ANP matrices for carbon with an effective aborption of $15\%$, $20\%$, and $25\%$  
reflecting the uncertainty of this quantity. The band of these three curves has to be compared to the data points.

\subsection{ CC$1\pi^+$ and CC$1\pi^0$ production}

\begin{figure}[thp]
\centering
\includegraphics[width=0.45\textwidth]{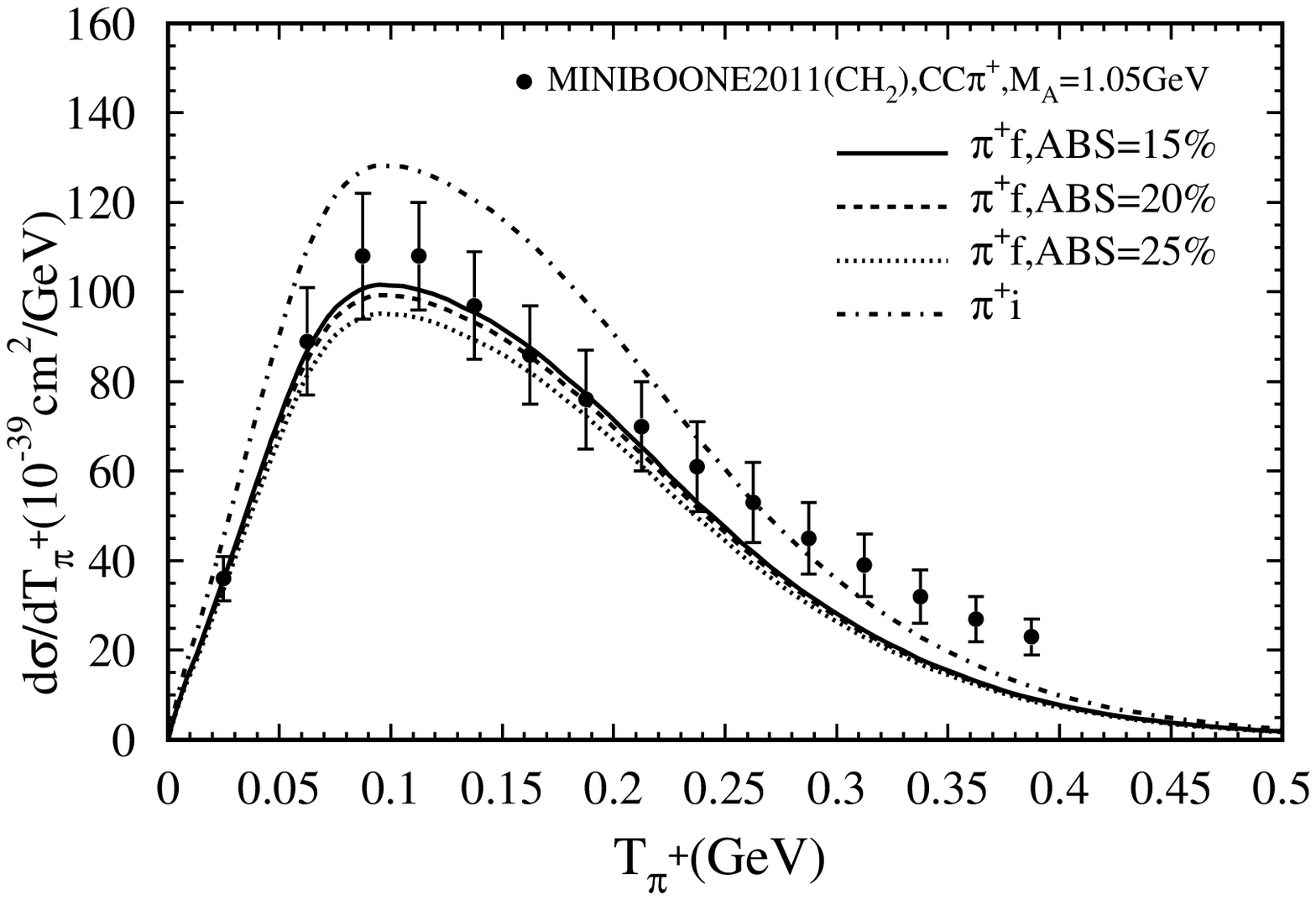}
\vspace*{-1.cm}
\caption{\sf 
Differential cross sections for CC$1\pi^+$ production in mineral oil
in dependence of the kinetic energy of the pion $T_\pi^+$
[cf.\ Tab.\ VI (Fig.\ 23) in \protect\cite{AguilarArevalo:2010bm}].
}
\label{fig:sigtpi}
\end{figure}

\begin{figure}[thp]
\centering
\includegraphics[width=0.45\textwidth]{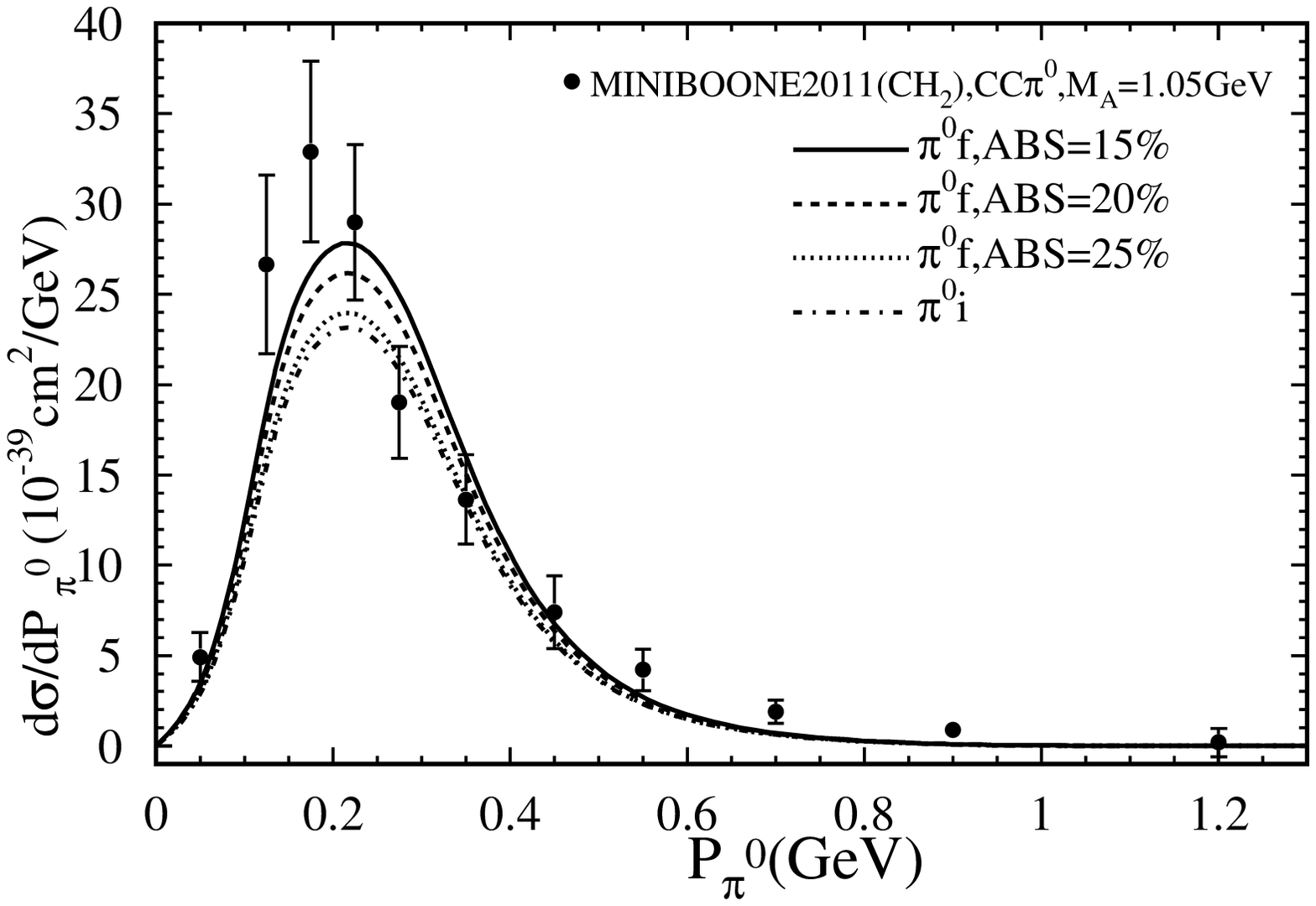}
\vspace*{-1.cm}
\caption{\sf 
Differential cross sections for CC$1\pi^0$ production in mineral oil
in dependence of the pion momentum $P_\pi^0$ 
[cf.\ Tab.\ X (Fig.\ 12) in \protect\cite{AguilarArevalo:2010xt}].
}
\label{fig:sigppi}
\end{figure}

We begin with our predictions for charged current single pion production
in the original framework using the input parameters given in Table \ref{tab:input}
and compare them with the MiniBooNE data from Ref.\ \cite{AguilarArevalo:2010bm} (CC$1\pi^+$)
and  Ref.\ \cite{AguilarArevalo:2010xt} (CC$1\pi^0$).

In Fig.\ \ref{fig:sigtot}, we show the unfolded total cross sections for CC$1\pi^+$ production (left) 
and CC$1\pi^0$ production (right) in dependence of the neutrino energy.
For energies $E_\nu \lesssim 1$ GeV we find perfect agreement between our cross sections
for neutral pion production and the data (right). Our results for the charged pion production (left)
are at the upper end of the very precise data at low neutrino energies ($E_\nu \lesssim 0.7$ GeV) 
but also in this case the overall description is very good.
For $E_\nu \gtrsim 1.4$ GeV our curves are systematically below the data by
roughly $1\sigma$ and with a flat energy dependence as expected for the $\Delta$ resonance.
At these higher energies the contributions from higher resonances and from deep inelastic scattering 
are expected to set in which are not included in our calculation and which could explain the discrepancy.
 
Results for the flux-averaged $Q^2$-differential cross sections are presented in Fig.\ \ref{fig:sigq2}
and as can be seen, the predicted $Q^2$-spectra for CC$1\pi^+$ (left) and CC$1\pi^0$ (right) 
slightly undershoot the data, in particular at larger $Q^2$.
As mentioned already, we don't neglect the muon mass in the calculation which has
a visible effect in the small $Q^2$ region. Setting the muon mass to zero would lead to a better
description of the data in the peak region and more significantly overshoot the data point in the lowest $Q^2$ bin.
It should be noted that the small $Q^2$ region has also been studied recently employing the ANP model
in Ref.\ \cite{Paschos:2012tr}. 

The comparison with the muon kinetic energy spectra is performed in Fig.\ \ref{fig:sigtmu}.
Again, the overall description of the data is not bad but our theoretical predictions slightly underestimate
them in the peak region. This is more pronounced in the case of neutral pion production.
Note that small $T_\mu$ correspond to large values of $W$ where higher resonances and/or a background
are more important and will move the theoretical curves higher once such contributions are included.

In Figs.\ \ref{fig:sigtpi} and \ref{fig:sigppi}, we show the differential cross sections for CC$1\pi^+$ production
in dependence of the kinetic energy of the pion and  for  CC$1\pi^0$ production in dependence of the pion 
momentum, respectively.
Similar to the previous figures, our theoretical curves are a bit low. 
In addition, in both Fig.\ \ref{fig:sigtpi} and \ref{fig:sigppi}, one can observe that our predicted cross sections 
are slightly harder than the data. 
This is better visible in Fig.\ \ref{fig:sigppi} due to the narrower spectrum.
Here the theory curves peak at $P_{\pi^0} \sim 230$ MeV whereas the MiniBooNE data have a peak at about 200 MeV.

\begin{figure}[t]
\centering
\includegraphics[width=0.45\textwidth]{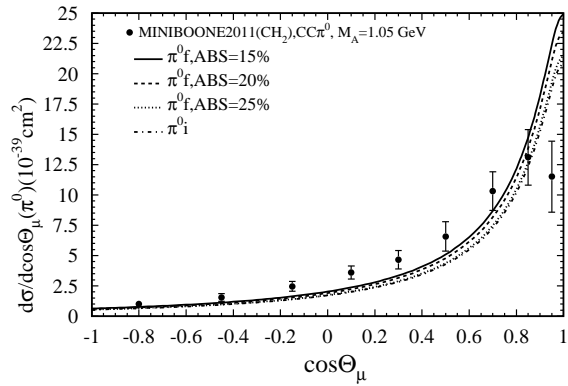}
\vspace*{-1.0cm}
\caption{\sf Differential cross sections for CC$1\pi^0$ production in mineral oil 
in dependence of the muon polar angle $\cos \theta_{\mu}$ 
[cf.\ Tab.\ IX (Fig.\ 11) in \protect\cite{AguilarArevalo:2010xt}].
} 
\label{fig:sigcosmu}
\end{figure}

\begin{figure}[t]
\centering
\includegraphics[width=0.45\textwidth]{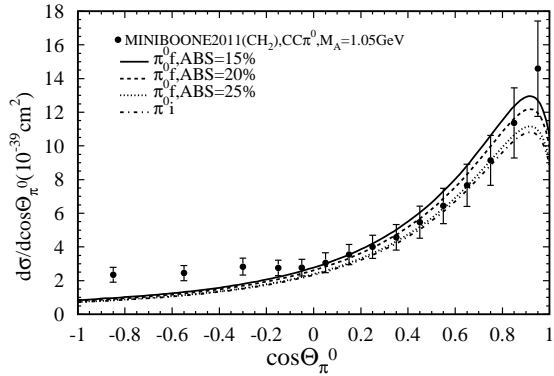}
\vspace*{-1.0cm}
\caption{\sf Same as in Fig.\ \protect\ref{fig:sigcosmu} for the pion polar angle $\cos \theta_{\pi^0}$ 
[cf.\ Tab.\ XI (Fig.\ 13) in \protect\cite{AguilarArevalo:2010xt}].
}
\label{fig:sigctpi0}
\end{figure}

\begin{figure*}[thp!]
\centering
\includegraphics[width=0.45\textwidth]{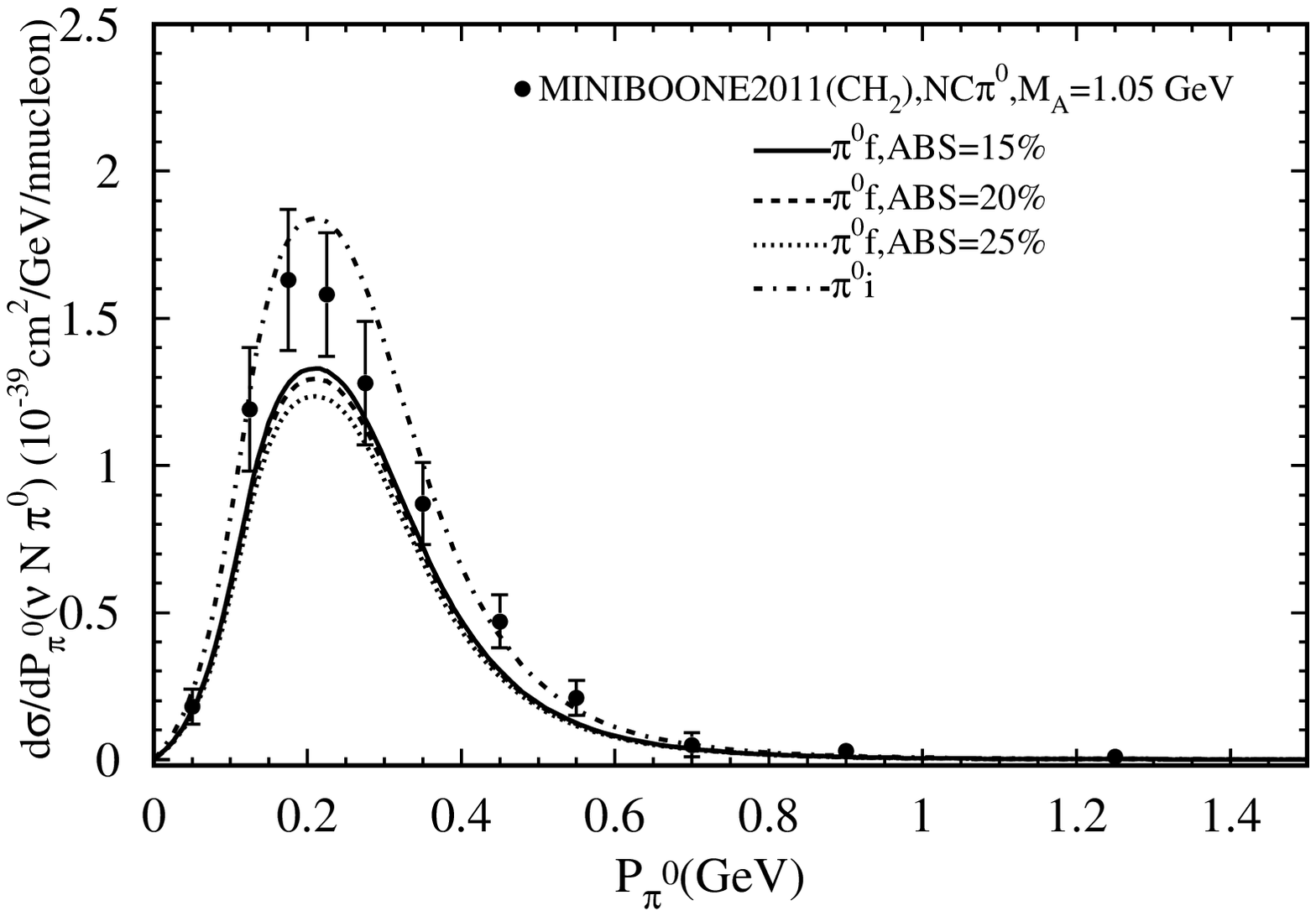}
\includegraphics[width=0.45\textwidth]{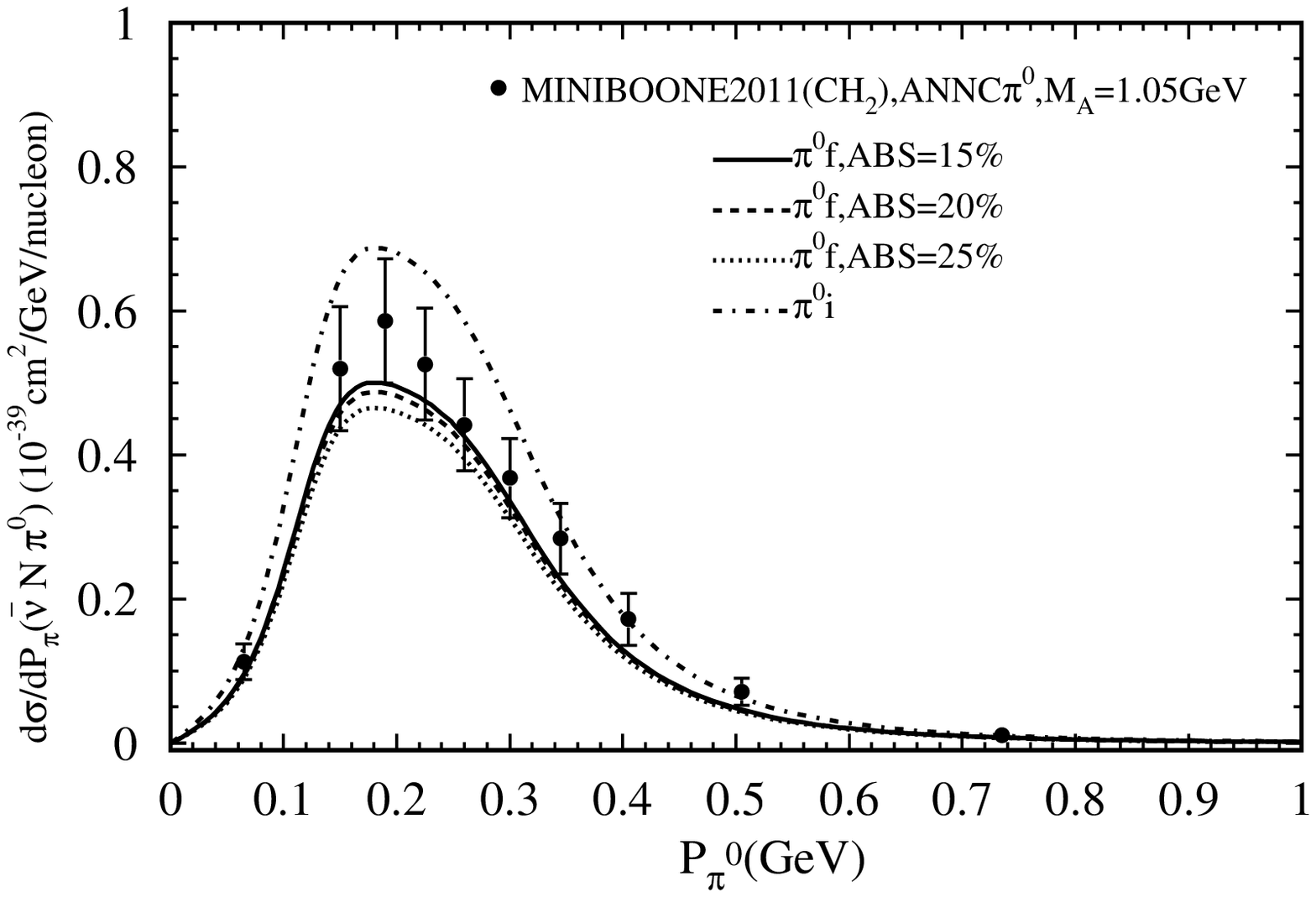}
\vspace*{-1.0cm}
\caption{\sf Differential cross sections for NC$1\pi^0$ production 
in $\nu_\mu$ scattering (left) and $\bar\nu_\mu$ scattering (right)
off a mineral oil target in dependence of the pion momentum
[cf.\ Tab.\ IV a), c) and Fig.\ 7 a), c) in \protect \cite{AguilarArevalo:2009ww}].
}
\label{fig:ncppi0}
\end{figure*}

\begin{figure*}[thp!]
\centering
\includegraphics[width=0.45\textwidth]{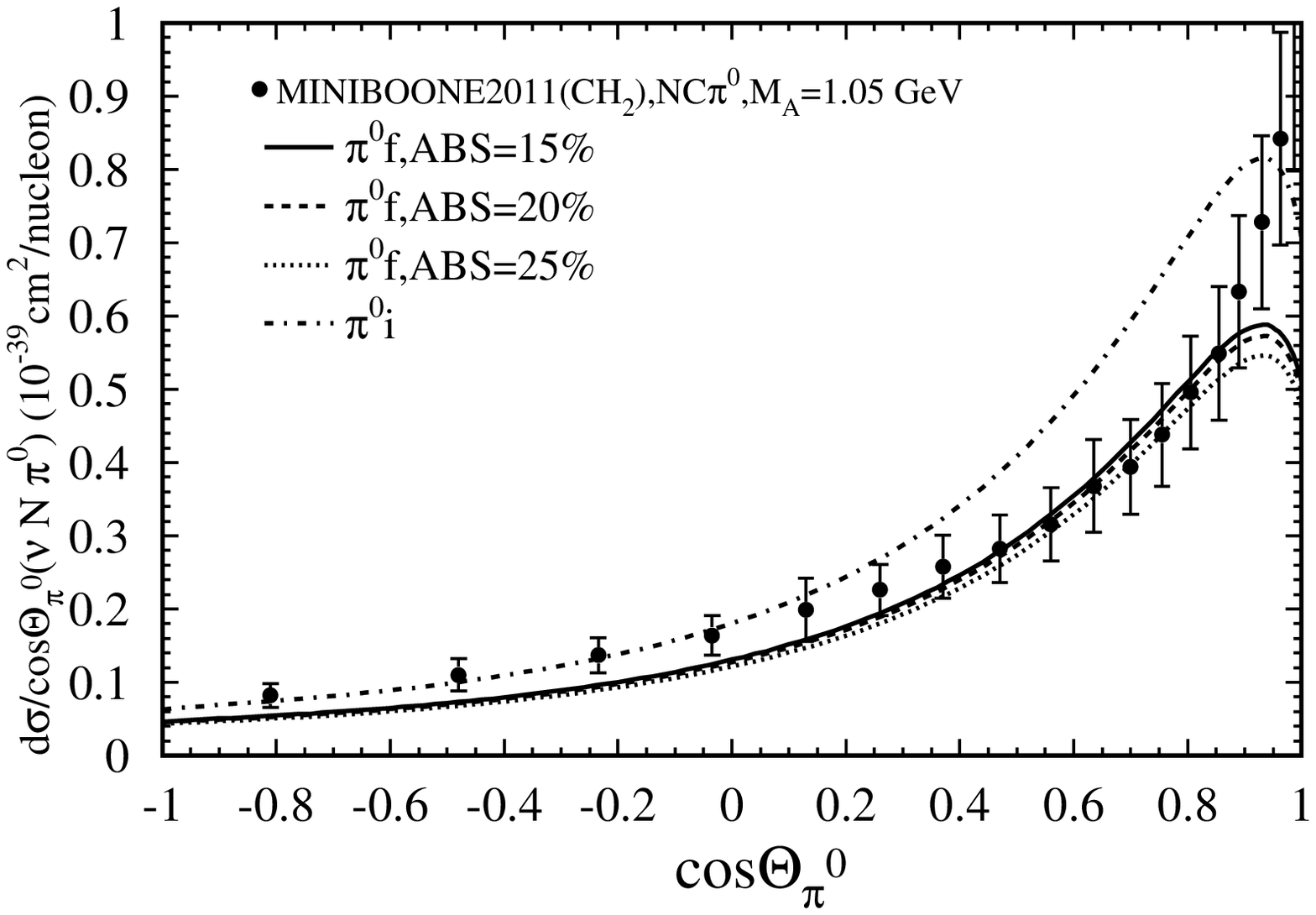}
\includegraphics[width=0.45\textwidth]{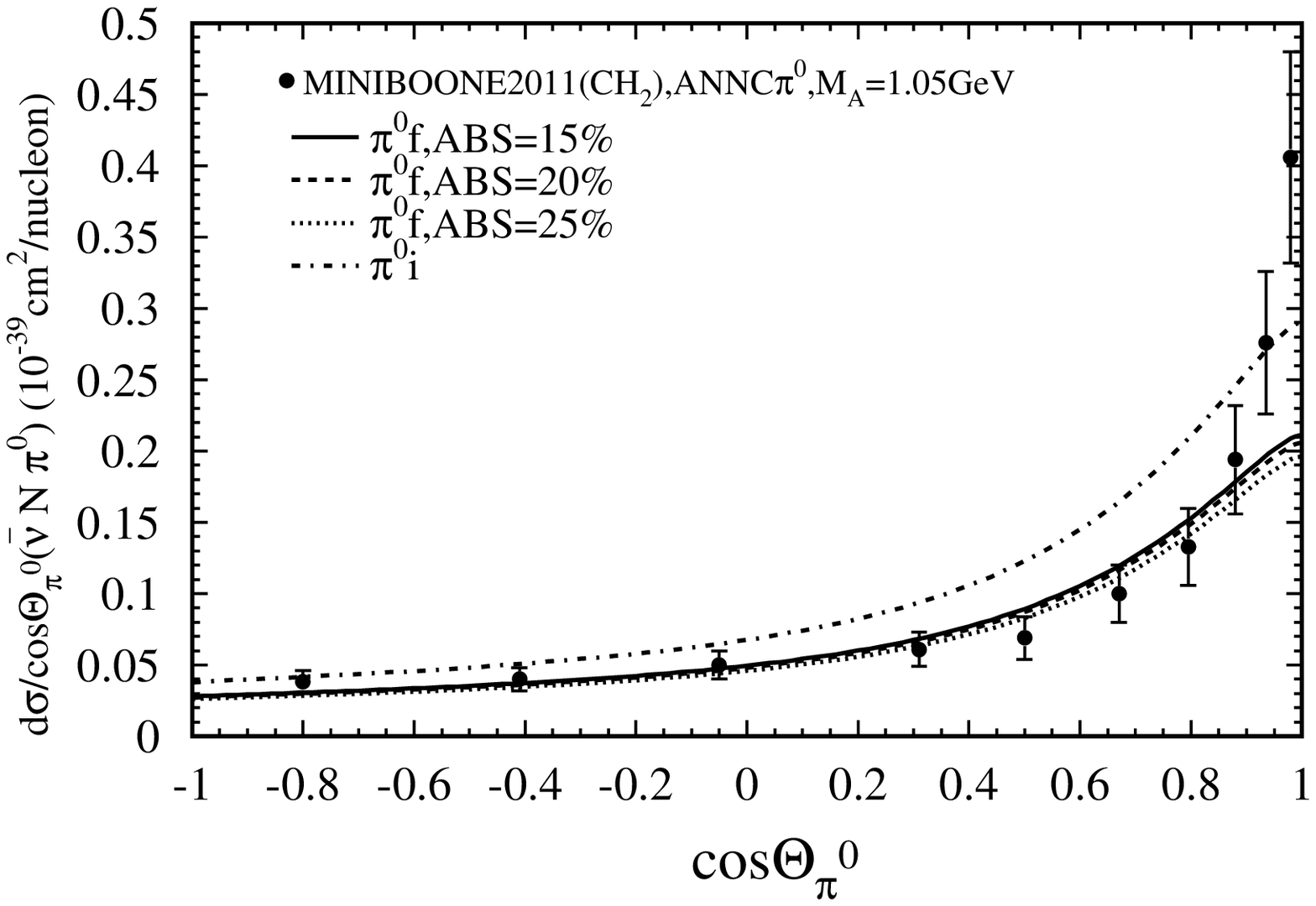}
\vspace*{-1.0cm}
\caption{\sf Same as in Fig.\ \protect\ref{fig:ncppi0} for the distributions in the pion polar angle
[cf.\ Tab.\ IV b), d) and Fig.\ 7 b), d) in \protect \cite{AguilarArevalo:2009ww}].
}
\label{fig:nccpi0}
\end{figure*}

Finally, we perform the comparison with the  angular distributions of CC$1\pi^0$ events.
The dependence of the differential cross section on the polar angle of the muon, $\cos\theta_\mu$, 
is presented in Fig.\ \ref{fig:sigcosmu}.
Our curves undershoot the data in the region $\cos \theta_\mu \in [-0.3, 0.4]$ which is
most significant in the central region where the data are more precise than in the forward region.
It should also be noted that the forward region $\cos \theta_\mu \to 1$ is correlated to the small $Q^2$ region
in Fig.\ \ref{fig:sigq2} (right).
The corresponding distribution in the polar angle of the pion in Fig.\ \ref{fig:sigctpi0} describes the data reasonably well in the forward region 
but clearly undershoots them in the backward region.

We can observe in Figs.\ \ref{fig:sigtot} -- \ref{fig:sigtmu}
that the cross sections for $\pi_f^+$ production are considerably smaller than the
free nucleon cross section $\pi^+_i$. Conversely, the cross sections for neutral pion production, $\pi^0_f$,
are of similar size as the free nucleon cross sections $\pi^0_i$
or even slightly enhanced in the cases of $15\%$ and $20\%$ effective absorption.
This is a generic feature of the ANP model which differs from other models in the literature 
\cite{Lalakulich:2012cj,Hernandez:2013jka}:
the larger cross sections (here the ones for CC$1\pi^+$ production) get reduced by both, the charge exchange effects
and the absorption, whereas for the smaller cross sections (here the ones for CC$1\pi^0$) the reduction due to the absorption is
(over-)compensated by an enhancement due to the charge exchange. 

It is also noteworthy that the predictions for CC$1\pi^0$ production show some dependence on 
the pion absorption in kinematic regions where the cross section is peaking.
On the other hand, the cross sections for CC$1\pi^+$ and NC$1\pi^0$ production (see below) are quite insensitive 
to the precise amount of pion absorption.

\subsection{NC$1\pi^0$ production in $\nu_\mu$ and $\bar\nu_\mu$ scattering}

We now turn to the discussion of the NC neutral pion production 
in $\nu_\mu$ ($\nu_\mu$-NC$1\pi^0$) and $\bar\nu_\mu$ ($\bar\nu_\mu$-NC$1\pi^0$) scattering.
We present predictions for flux-averaged cross sections in mineral oil
and compare them to the data reported by the MiniBooNE collaboration in \cite{AguilarArevalo:2009ww}
from where we also take the fluxes for the neutrinos and anti-neutrinos.

In Fig.\ \ref{fig:ncppi0}, we show results for $\nu_\mu$-NC$1\pi^0$ production (left) and $\bar\nu_\mu$-NC$1\pi^0$ production (right)
in dependence of the pion momentum and find that the shape of the data is nicely described by our
predictions with a slightly too small normalization.
Different from Fig.\ \ref{fig:sigppi}, the peak positions of the data and the theoretical predictions are consistent.

Finally, the distributions in the pion polar angle are presented in Fig.\ \ref{fig:nccpi0} for $\nu_\mu$-scattering (left) and
$\bar\nu_\mu$-scattering (right).
As can be seen, the overall agreement with the data is better than in the charged current $\pi^0$ case (see Fig.\ \ref{fig:sigctpi0}).
In the case of $\bar\nu_\mu$-scattering (right figure) the description of the data is even very good except in the very
forward region where our curves are below the data.
This difference may be accounted for by the small positive contribution of the coherent pion production cross sections.
It is very interesting to note that the excess of the NC1$\pi^0$ data at small $\theta_\pi$ appears to be equal for the
neutrino and anti-neutrino scattering as predicted by theory \cite{Paschos:2012tr}.
Our curves can be used for subtracting the background from resonance production and thus estimating the coherent cross section.
We shall return to this topic and fit the NC1$\pi^0$ data including coherent scattering in a future publication.

In summary, we have compared our original theoretical predictions to the complete set of available data from MiniBooNE.
The overall description of the data is acceptable in particular concerning the shapes.
However, in general, the normalization of the theory curves is too small as has also been observed by other groups 
\cite{Lalakulich:2012cj,Hernandez:2013jka}.
This discrepancy might be explained by missing contributions from a non-resonant background and/or higher resonances
despite the experimental cut $W<1.35$ GeV which has been used in the analysis of the CC1$\pi^+$ events. 


\section{Comparison with MINER$\nu$A data}
\label{sec:minerva}

In this section, we present a comparison of our cross section predictions with the most recent data
on single charged pion production from the MINER$\nu$A collaboration \cite{Eberly:2014mra}.
The MINER$\nu$A experiment is exposed to the NuMI wideband neutrino beam at Fermilab. 
The neutrino beam is higher compared to the MiniBooNe experiment and the events have energies in the range 
$1.5 < E_\nu < 10$ GeV and the average energy is $\langle E_\nu \rangle = 4.0$ GeV.
In addition, the target is plastic (CH) and thus contains one Hydrogen atom less than the mineral oil (CH$_2$) of MiniBooNE.
Furthermore, to isolate the contribution from single pion production a cut on the invariant mass $W<1.4$ GeV has been applied
in the experimental analysis and in our calculation.

\begin{figure}[t]
\centering
\includegraphics[width=0.45\textwidth]{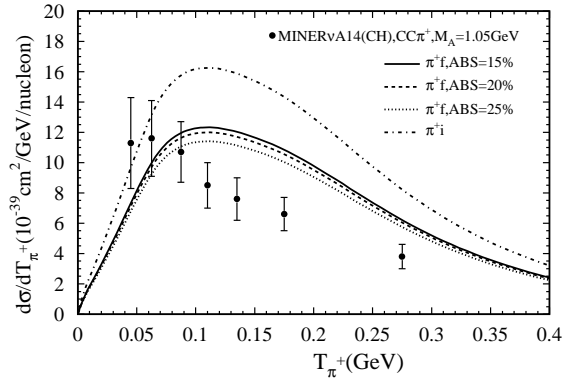}
\vspace*{-1.0cm}
\caption{\sf Flux-averaged differential cross sections per nucleon for CC$1\pi^+$ production in plastic (CH)
in dependence of the kinetic energy of the pion $T_\pi^+$ in comparison with MINER$\nu$A data \protect\cite{Eberly:2014mra}.
}
\label{fig:minervatpi}
\end{figure}

The results for the flux-averaged differential cross section $d\sigma/dT_\pi$ as function of the pion kinetic energy $T_\pi$ are
presented in Fig.\ \ref{fig:minervatpi}.
Our predictions show a prominent albeit rather broad peak at $T_\pi \sim 120$ MeV which is not reflected by the MINER$\nu$A data. 
However, a peak in the data may be located at a lower value of $T_\pi \sim 70$ MeV. This is to be compared to the MiniBooNE data with a
peak at $T_\pi \sim 100$ MeV  which is also consistent with the maximum of the pion momentum distributions at $P_\pi \sim 200$ MeV.
While the general normalization is reproduced our results undershoot (overshoot) the data by 
about 1$\sigma$ for $T_\pi < 70$ MeV ($T_\pi > 150$ MeV).
The final state interactions in the ANP model due to pion absorption and charge exchange have a noticeable effect
on the cross section and lead to an improved description of the data. 
The fact, that the pion energy is not modified in the ANP model (unless the pion is absorbed)
might explain why the theoretical curves appear to be right-shifted by about 50 MeV with respect to the data.
Clearly, pion energy loss effects would lead to a softer spectrum.

\begin{figure}[t]
\centering
\includegraphics[width=0.45\textwidth]{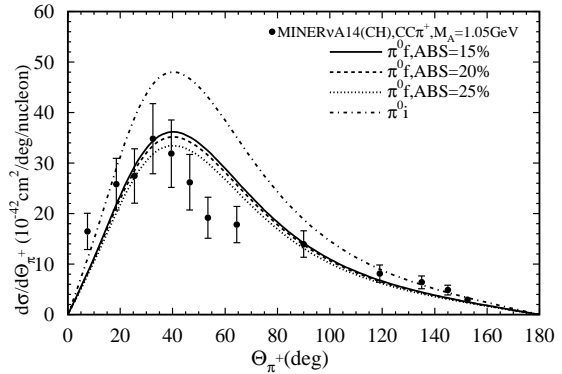}
\vspace*{-1.0cm}
\caption{\sf  Flux-averaged CC$1\pi^+$ production cross sections per nucleon in plastic (CH) as function of the pion angle 
in comparison with MINER$\nu$A data \protect\cite{Eberly:2014mra}. 
}
\label{fig:thetapi}
\end{figure}

We also computed the differential cross section on the pion angle relative to the beam direction.
The results in absolute units are shown in Fig.\ \ref{fig:thetapi}.
Since the data were presented in arbitrary units we have normalized them to our theory prediction.
As can be seen, the measured shape is well reproduced by our model.

\section{Summary and conclusions}
\label{sec:summary}

The article presents theoretical predictions for single pion production ($\pi^+, \pi^0$)
in neutrino--nucleus scattering in the framework of an earlier model
and performs a comprehensive comparison with all the experimental results on single pion production from MiniBooNE 
and MINER$\nu$A in order to obtain a complete picture of the situation.
The model includes three resonances ($\Delta$, $P_{11}$ and $S_{11}$) folded with the nuclear corrections of the ANP model
which accounts for the intranuclear rescattering effects of the final state pions.
We do not include contributions from a non-resonant background, coherent scattering and deep inelastic scattering.

In general the theoretical curves reproduce the shape of the  MiniBooNe data, but in certain regions of phase space 
they are below the data by 1 or 2$\sigma$. The agreement with the neutral current data is slightly better.

The fact that the measured integrated cross section rises with energy indicates that 
in our calculation there are missing contributions from higher resonances and the production of the continuum 
despite of the cut on the invariant mass $W<1.35$ GeV which has been
used in the case of CC1$\pi^+$ events whose aim is to suppress such contributions.
Clearly, this difference between the measured and the predicted total cross section at higher neutrino energies is also
reflected in the differential distributions.
Larger discrepancies are observed at $Q^2 \to 0$ (and, equivalently, $\cos \theta_\mu \to 1$), small $T_\mu$ (which is
correlated with large invariant masses $W$), and in the backward region of the pion polar angle.
In the small $Q^2$ region progress has been made in Ref.\ \cite{Paschos:2012tr} and these results can be 
used to improve our theory in the future.

\begin{figure}[th]
\centering
\includegraphics[width=0.45\textwidth]{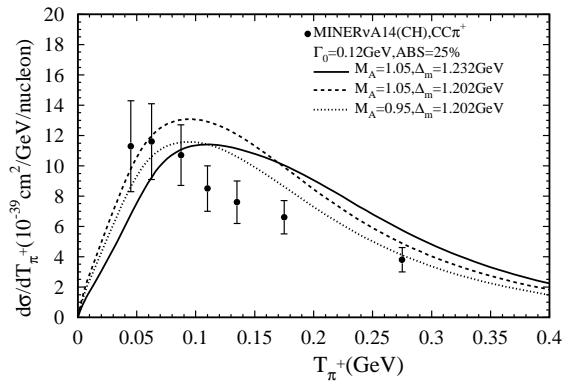}
\vspace*{-1.0cm}
\caption{\sf Comparison of theoretical predictions using different values for the axial mass and the mass of the $\Delta$ resonance with
the MINER$\nu$A data for the distribution in the kinetic energy of the pion \protect\cite{Eberly:2014mra}.
The solid line shows the curve with $M_A=1.05$ GeV and $M_\Delta=1.232$ GeV already presented
in Fig.\ \protect\ref{fig:minervatpi}.
In addition we show results for $M_A=1.05$ GeV, $M_\Delta=1.202$ GeV (dashed line) and
$M_A=0.95$ GeV, $M_\Delta=1.202$ GeV (dotted line).
}
\label{fig:minervatpi2}
\end{figure}

Despite the fact that the energy spectrum of the neutrino beam is higher in the MINER$\nu$A experiment permitting additional contributions
from higher energies, the normalization of our theoretical curves is in good agreement with the MINER$\nu$A
data for the kinetic energy of the pion, presumably due to the cut on the invariant mass of the produced hadronic system $W<1.4$ GeV. 
Furthermore, the predicted angular dependence on the polar angle of the pion is in good agreement with the observed shape.
However, the shape of the $T_\pi$ spectrum measured by MINER$\nu$A is softer compared to the theory curves.
It doesn't have a peak at $T_\pi \sim 100$ MeV as seen in the corresponding MiniBooNE data (Fig.\ \ref{fig:sigtpi}). 
Possibly the maximum is located at $T_\pi \sim 70$ MeV but this is not fully clear.
Such a shift to lower energies could be explained by energy loss effects during the propagation of the pion
through the nucleus not taken into account in the ANP model.
An alternative/additional explanation could be in-medium modifications of single pion production caused by a change of the mass and the width
of the $\Delta$ resonance in the nuclear environment \cite{Lalakulich:2012cj,Hernandez:2013jka}.
To illustrate this effect we show in Fig.\ \ref{fig:minervatpi2} a comparison of our original prediction (solid line) for the 
$T_\pi$ spectrum with results obtained with modified parameters in step 1 of our calculation.
As expected, a smaller $M_\Delta = 1.202$ GeV (dashed line) leads to a softer spectrum with an enhanced
cross section at the peak. Yet a smaller $M_\Delta$ would be needed for a curve with a maximum at $T_\pi\sim 70$ MeV.
The best description of the data is obtained using a reduced $\Delta$ mass together with a smaller axial vector
mass $M_A=0.95$ GeV in order to compensate for the enhancement of the cross section at the peak position.

We have performed more detailed studies of the effect of medium-modified parameters ($M_A, M_\Delta, \Gamma_0$).
However, we refrain from showing them in this article since a clear preference did not emerge.
First of all, while the distribution in the pion momentum in Fig.\ \ref{fig:sigppi} is also shifted
to the left, this is not the case in the corresponding Fig.\ \ref{fig:ncppi0}.
Second, we have not found a set of parameters which would improve the overall description of the
complete data set.

At the moment it is hard to decide whether we should add a background or if we should modify the parameters 
(mass, width and/or form factors) of the $\Delta$ resonance or incorporate a mechanism for pion energy loss into
the ANP formalism.
To get a clearer picture, it would be useful to have detailed information on the $W$ distributions
for different nuclear targets. The results for free nucleon and light nuclear targets should be compared to our Fig.\ \ref{fig:sigw1}.
Furthermore, separate measurements of the $W$ distributions for charged and neutral pions (in the region of the $\Delta$ resonance) 
would provide excellent tests of the charge exchange and absorption effects as predicted by the ANP model \cite{Paschos:2007pe}.
Finally, a separation of the coherent pion production would be useful.
%


\section*{Acknowledgments}
We are grateful to Jorge Morfin and Brandon Eberly for useful discussions and comments on an earlier version of the 
manuscript of the paper.

\clearpage

\bibliographystyle{apsrev4-1}
\bibliography{single_pion}

\begin{thebibliography}{28}%
\makeatletter
\providecommand \@ifxundefined [1]{%
 \@ifx{#1\undefined}
}%
\providecommand \@ifnum [1]{%
 \ifnum #1\expandafter \@firstoftwo
 \else \expandafter \@secondoftwo
 \fi
}%
\providecommand \@ifx [1]{%
 \ifx #1\expandafter \@firstoftwo
 \else \expandafter \@secondoftwo
 \fi
}%
\providecommand \natexlab [1]{#1}%
\providecommand \enquote  [1]{``#1''}%
\providecommand \bibnamefont  [1]{#1}%
\providecommand \bibfnamefont [1]{#1}%
\providecommand \citenamefont [1]{#1}%
\providecommand \href@noop [0]{\@secondoftwo}%
\providecommand \href [0]{\begingroup \@sanitize@url \@href}%
\providecommand \@href[1]{\@@startlink{#1}\@@href}%
\providecommand \@@href[1]{\endgroup#1\@@endlink}%
\providecommand \@sanitize@url [0]{\catcode `\\12\catcode `\$12\catcode
  `\&12\catcode `\#12\catcode `\^12\catcode `\_12\catcode `\%12\relax}%
\providecommand \@@startlink[1]{}%
\providecommand \@@endlink[0]{}%
\providecommand \url  [0]{\begingroup\@sanitize@url \@url }%
\providecommand \@url [1]{\endgroup\@href {#1}{\urlprefix }}%
\providecommand \urlprefix  [0]{URL }%
\providecommand \Eprint [0]{\href }%
\providecommand \doibase [0]{http://dx.doi.org/}%
\providecommand \selectlanguage [0]{\@gobble}%
\providecommand \bibinfo  [0]{\@secondoftwo}%
\providecommand \bibfield  [0]{\@secondoftwo}%
\providecommand \translation [1]{[#1]}%
\providecommand \BibitemOpen [0]{}%
\providecommand \bibitemStop [0]{}%
\providecommand \bibitemNoStop [0]{.\EOS\space}%
\providecommand \EOS [0]{\spacefactor3000\relax}%
\providecommand \BibitemShut  [1]{\csname bibitem#1\endcsname}%
\let\auto@bib@innerbib\@empty
\bibitem [{\citenamefont {Morfin}\ \emph {et~al.}(2012)\citenamefont {Morfin},
  \citenamefont {Nieves},\ and\ \citenamefont {Sobczyk}}]{Morfin:2012kn}%
  \BibitemOpen
  \bibfield  {author} {\bibinfo {author} {\bibfnamefont {J.~G.}\ \bibnamefont
  {Morfin}}, \bibinfo {author} {\bibfnamefont {J.}~\bibnamefont {Nieves}}, \
  and\ \bibinfo {author} {\bibfnamefont {J.~T.}\ \bibnamefont {Sobczyk}},\
  }\href {\doibase 10.1155/2012/934597} {\bibfield  {journal} {\bibinfo
  {journal} {Adv.High Energy Phys.}\ }\textbf {\bibinfo {volume} {2012}},\
  \bibinfo {pages} {934597} (\bibinfo {year} {2012})},\ \Eprint
  {http://arxiv.org/abs/1209.6586} {arXiv:1209.6586 [hep-ex]} \BibitemShut
  {NoStop}%
\bibitem [{\citenamefont {Aguilar-Arevalo}\ \emph
  {et~al.}(2011{\natexlab{a}})\citenamefont {Aguilar-Arevalo} \emph
  {et~al.}}]{AguilarArevalo:2010bm}%
  \BibitemOpen
  \bibfield  {author} {\bibinfo {author} {\bibfnamefont {A.}~\bibnamefont
  {Aguilar-Arevalo}} \emph {et~al.} (\bibinfo {collaboration} {MiniBooNE
  Collaboration}),\ }\href {\doibase 10.1103/PhysRevD.83.052007} {\bibfield
  {journal} {\bibinfo  {journal} {Phys.Rev.}\ }\textbf {\bibinfo {volume}
  {D83}},\ \bibinfo {pages} {052007} (\bibinfo {year} {2011}{\natexlab{a}})},\
  \Eprint {http://arxiv.org/abs/1011.3572} {arXiv:1011.3572 [hep-ex]}
  \BibitemShut {NoStop}%
\bibitem [{\citenamefont {Aguilar-Arevalo}\ \emph
  {et~al.}(2011{\natexlab{b}})\citenamefont {Aguilar-Arevalo} \emph
  {et~al.}}]{AguilarArevalo:2010xt}%
  \BibitemOpen
  \bibfield  {author} {\bibinfo {author} {\bibfnamefont {A.}~\bibnamefont
  {Aguilar-Arevalo}} \emph {et~al.} (\bibinfo {collaboration} {MiniBooNE
  Collaboration}),\ }\href {\doibase 10.1103/PhysRevD.83.052009} {\bibfield
  {journal} {\bibinfo  {journal} {Phys.Rev.}\ }\textbf {\bibinfo {volume}
  {D83}},\ \bibinfo {pages} {052009} (\bibinfo {year} {2011}{\natexlab{b}})},\
  \Eprint {http://arxiv.org/abs/1010.3264} {arXiv:1010.3264 [hep-ex]}
  \BibitemShut {NoStop}%
\bibitem [{\citenamefont {Aguilar-Arevalo}\ \emph {et~al.}(2010)\citenamefont
  {Aguilar-Arevalo} \emph {et~al.}}]{AguilarArevalo:2009ww}%
  \BibitemOpen
  \bibfield  {author} {\bibinfo {author} {\bibfnamefont {A.~A.}\ \bibnamefont
  {Aguilar-Arevalo}} \emph {et~al.} (\bibinfo {collaboration} {MiniBooNE
  Collaboration}),\ }\href {\doibase 10.1103/PhysRevD.81.013005} {\bibfield
  {journal} {\bibinfo  {journal} {Phys.Rev.}\ }\textbf {\bibinfo {volume}
  {D81}},\ \bibinfo {pages} {013005} (\bibinfo {year} {2010})},\ \Eprint
  {http://arxiv.org/abs/0911.2063} {arXiv:0911.2063 [hep-ex]} \BibitemShut
  {NoStop}%
\bibitem [{\citenamefont {Lalakulich}\ and\ \citenamefont
  {Mosel}(2013)}]{Lalakulich:2012cj}%
  \BibitemOpen
  \bibfield  {author} {\bibinfo {author} {\bibfnamefont {O.}~\bibnamefont
  {Lalakulich}}\ and\ \bibinfo {author} {\bibfnamefont {U.}~\bibnamefont
  {Mosel}},\ }\href {\doibase 10.1103/PhysRevC.87.014602} {\bibfield  {journal}
  {\bibinfo  {journal} {Phys.Rev.}\ }\textbf {\bibinfo {volume} {C87}},\
  \bibinfo {pages} {014602} (\bibinfo {year} {2013})},\ \Eprint
  {http://arxiv.org/abs/1210.4717} {arXiv:1210.4717 [nucl-th]} \BibitemShut
  {NoStop}%
\bibitem [{\citenamefont {Hern{\'a}ndez}\ \emph {et~al.}(2013)\citenamefont
  {Hern{\'a}ndez}, \citenamefont {Nieves},\ and\ \citenamefont
  {Vacas}}]{Hernandez:2013jka}%
  \BibitemOpen
  \bibfield  {author} {\bibinfo {author} {\bibfnamefont {E.}~\bibnamefont
  {Hern{\'a}ndez}}, \bibinfo {author} {\bibfnamefont {J.}~\bibnamefont
  {Nieves}}, \ and\ \bibinfo {author} {\bibfnamefont {M.~J.~V.}\ \bibnamefont
  {Vacas}},\ }\href {\doibase 10.1103/PhysRevD.87.113009} {\bibfield  {journal}
  {\bibinfo  {journal} {Phys.Rev.}\ }\textbf {\bibinfo {volume} {D87}},\
  \bibinfo {pages} {113009} (\bibinfo {year} {2013})},\ \Eprint
  {http://arxiv.org/abs/1304.1320} {arXiv:1304.1320 [hep-ph]} \BibitemShut
  {NoStop}%
\bibitem [{\citenamefont {{Paschos, E. A. and Schalla,
  D.}}(2013)}]{Paschos:2012tr}%
  \BibitemOpen
  \bibfield  {author} {\bibinfo {author} {\bibnamefont {{Paschos, E. A. and
  Schalla, D.}}},\ }\href@noop {} {\bibfield  {journal} {\bibinfo  {journal}
  {Advances in High Energy Physics}\ }\textbf {\bibinfo {volume} {2013}},\
  \bibinfo {pages} {270792} (\bibinfo {year} {2013})},\ \Eprint
  {http://arxiv.org/abs/1212.4662} {arXiv:1212.4662} \BibitemShut {NoStop}%
\bibitem [{\citenamefont {Paschos}\ \emph {et~al.}(2000)\citenamefont
  {Paschos}, \citenamefont {Pasquali},\ and\ \citenamefont
  {Yu}}]{Paschos:2000be}%
  \BibitemOpen
  \bibfield  {author} {\bibinfo {author} {\bibfnamefont {E.~A.}\ \bibnamefont
  {Paschos}}, \bibinfo {author} {\bibfnamefont {L.}~\bibnamefont {Pasquali}}, \
  and\ \bibinfo {author} {\bibfnamefont {J.-Y.}\ \bibnamefont {Yu}},\ }\href
  {\doibase 10.1016/S0550-3213(00)00486-7} {\bibfield  {journal} {\bibinfo
  {journal} {Nucl.Phys.}\ }\textbf {\bibinfo {volume} {B588}},\ \bibinfo
  {pages} {263} (\bibinfo {year} {2000})},\ \Eprint
  {http://arxiv.org/abs/hep-ph/0005255} {arXiv:hep-ph/0005255 [hep-ph]}
  \BibitemShut {NoStop}%
\bibitem [{\citenamefont {Schienbein}\ and\ \citenamefont
  {Yu}(2003)}]{Schienbein:2003sm}%
  \BibitemOpen
  \bibfield  {author} {\bibinfo {author} {\bibfnamefont {I.}~\bibnamefont
  {Schienbein}}\ and\ \bibinfo {author} {\bibfnamefont {J.-Y.}\ \bibnamefont
  {Yu}},\ }\href@noop {} {\  (\bibinfo {year} {2003})},\ \Eprint
  {http://arxiv.org/abs/hep-ph/0308010} {arXiv:hep-ph/0308010 [hep-ph]}
  \BibitemShut {NoStop}%
\bibitem [{\citenamefont {Paschos}\ \emph {et~al.}(2004)\citenamefont
  {Paschos}, \citenamefont {Yu},\ and\ \citenamefont
  {Sakuda}}]{Paschos:2003qr}%
  \BibitemOpen
  \bibfield  {author} {\bibinfo {author} {\bibfnamefont {E.~A.}\ \bibnamefont
  {Paschos}}, \bibinfo {author} {\bibfnamefont {J.-Y.}\ \bibnamefont {Yu}}, \
  and\ \bibinfo {author} {\bibfnamefont {M.}~\bibnamefont {Sakuda}},\ }\href
  {\doibase 10.1103/PhysRevD.69.014013} {\bibfield  {journal} {\bibinfo
  {journal} {Phys.Rev.}\ }\textbf {\bibinfo {volume} {D69}},\ \bibinfo {pages}
  {014013} (\bibinfo {year} {2004})},\ \Eprint
  {http://arxiv.org/abs/hep-ph/0308130} {arXiv:hep-ph/0308130 [hep-ph]}
  \BibitemShut {NoStop}%
\bibitem [{\citenamefont {Paschos}\ \emph
  {et~al.}(2005{\natexlab{a}})\citenamefont {Paschos}, \citenamefont
  {Schienbein},\ and\ \citenamefont {Yu}}]{Paschos:2004qh}%
  \BibitemOpen
  \bibfield  {author} {\bibinfo {author} {\bibfnamefont {E.~A.}\ \bibnamefont
  {Paschos}}, \bibinfo {author} {\bibfnamefont {I.}~\bibnamefont {Schienbein}},
  \ and\ \bibinfo {author} {\bibfnamefont {J.-Y.}\ \bibnamefont {Yu}},\ }\href
  {\doibase 10.1016/j.nuclphysbps.2004.11.241} {\bibfield  {journal} {\bibinfo
  {journal} {Nucl.Phys.Proc.Suppl.}\ }\textbf {\bibinfo {volume} {139}},\
  \bibinfo {pages} {119} (\bibinfo {year} {2005}{\natexlab{a}})},\ \Eprint
  {http://arxiv.org/abs/hep-ph/0408148} {arXiv:hep-ph/0408148 [hep-ph]}
  \BibitemShut {NoStop}%
\bibitem [{\citenamefont {Paschos}\ \emph
  {et~al.}(2005{\natexlab{b}})\citenamefont {Paschos}, \citenamefont {Sakuda},
  \citenamefont {Schienbein},\ and\ \citenamefont {Yu}}]{Paschos:2004md}%
  \BibitemOpen
  \bibfield  {author} {\bibinfo {author} {\bibfnamefont {E.~A.}\ \bibnamefont
  {Paschos}}, \bibinfo {author} {\bibfnamefont {M.}~\bibnamefont {Sakuda}},
  \bibinfo {author} {\bibfnamefont {I.}~\bibnamefont {Schienbein}}, \ and\
  \bibinfo {author} {\bibfnamefont {J.-Y.}\ \bibnamefont {Yu}},\ }\href
  {\doibase 10.1016/j.nuclphysbps.2004.11.205} {\bibfield  {journal} {\bibinfo
  {journal} {Nucl.Phys.Proc.Suppl.}\ }\textbf {\bibinfo {volume} {139}},\
  \bibinfo {pages} {125} (\bibinfo {year} {2005}{\natexlab{b}})},\ \Eprint
  {http://arxiv.org/abs/hep-ph/0408185} {arXiv:hep-ph/0408185 [hep-ph]}
  \BibitemShut {NoStop}%
\bibitem [{\citenamefont {Eberly}\ \emph {et~al.}(2014)\citenamefont {Eberly}
  \emph {et~al.}}]{Eberly:2014mra}%
  \BibitemOpen
  \bibfield  {author} {\bibinfo {author} {\bibfnamefont {B.}~\bibnamefont
  {Eberly}} \emph {et~al.} (\bibinfo {collaboration} {The MINERvA
  Collaboration}),\ }\href@noop {} {\  (\bibinfo {year} {2014})},\ \Eprint
  {http://arxiv.org/abs/1406.6415} {arXiv:1406.6415 [hep-ex]} \BibitemShut
  {NoStop}%
\bibitem [{\citenamefont {Paschos}\ \emph {et~al.}()\citenamefont {Paschos},
  \citenamefont {Schienbein},\ and\ \citenamefont {Yu}}]{psy-winp}%
  \BibitemOpen
  \bibfield  {author} {\bibinfo {author} {\bibfnamefont {E.~A.}\ \bibnamefont
  {Paschos}}, \bibinfo {author} {\bibfnamefont {I.}~\bibnamefont {Schienbein}},
  \ and\ \bibinfo {author} {\bibfnamefont {J.~Y.}\ \bibnamefont {Yu}},\
  }\href@noop {} {}\bibinfo {note} {{work in progress}}\BibitemShut {NoStop}%
\bibitem [{\citenamefont {Fogli}\ and\ \citenamefont
  {Nardulli}(1979)}]{Fogli:1979cz}%
  \BibitemOpen
  \bibfield  {author} {\bibinfo {author} {\bibfnamefont {G.~L.}\ \bibnamefont
  {Fogli}}\ and\ \bibinfo {author} {\bibfnamefont {G.}~\bibnamefont
  {Nardulli}},\ }\href {\doibase 10.1016/0550-3213(79)90233-5} {\bibfield
  {journal} {\bibinfo  {journal} {Nucl.Phys.}\ }\textbf {\bibinfo {volume}
  {B160}},\ \bibinfo {pages} {116} (\bibinfo {year} {1979})}\BibitemShut
  {NoStop}%
\bibitem [{\citenamefont {Fogli}\ and\ \citenamefont
  {Nardulli}(1980)}]{Fogli:1979qj}%
  \BibitemOpen
  \bibfield  {author} {\bibinfo {author} {\bibfnamefont {G.~L.}\ \bibnamefont
  {Fogli}}\ and\ \bibinfo {author} {\bibfnamefont {G.}~\bibnamefont
  {Nardulli}},\ }\href {\doibase 10.1016/0550-3213(80)90312-0} {\bibfield
  {journal} {\bibinfo  {journal} {Nucl.Phys.}\ }\textbf {\bibinfo {volume}
  {B165}},\ \bibinfo {pages} {162} (\bibinfo {year} {1980})}\BibitemShut
  {NoStop}%
\bibitem [{\citenamefont {Schreiner}\ and\ \citenamefont
  {Von~Hippel}(1973)}]{Schreiner:1973mj}%
  \BibitemOpen
  \bibfield  {author} {\bibinfo {author} {\bibfnamefont {P.~A.}\ \bibnamefont
  {Schreiner}}\ and\ \bibinfo {author} {\bibfnamefont {F.}~\bibnamefont
  {Von~Hippel}},\ }\href {\doibase 10.1016/0550-3213(73)90588-9} {\bibfield
  {journal} {\bibinfo  {journal} {Nucl.Phys.}\ }\textbf {\bibinfo {volume}
  {B58}},\ \bibinfo {pages} {333} (\bibinfo {year} {1973})}\BibitemShut
  {NoStop}%
\bibitem [{\citenamefont {Rein}\ and\ \citenamefont
  {Sehgal}(1981)}]{Rein:1980wg}%
  \BibitemOpen
  \bibfield  {author} {\bibinfo {author} {\bibfnamefont {D.}~\bibnamefont
  {Rein}}\ and\ \bibinfo {author} {\bibfnamefont {L.~M.}\ \bibnamefont
  {Sehgal}},\ }\href {\doibase 10.1016/0003-4916(81)90242-6} {\bibfield
  {journal} {\bibinfo  {journal} {Annals Phys.}\ }\textbf {\bibinfo {volume}
  {133}},\ \bibinfo {pages} {79} (\bibinfo {year} {1981})}\BibitemShut
  {NoStop}%
\bibitem [{\citenamefont {Radecky}\ \emph {et~al.}(1982)\citenamefont
  {Radecky}, \citenamefont {Barnes}, \citenamefont {Carmony}, \citenamefont
  {Garfinkel}, \citenamefont {Derrick} \emph {et~al.}}]{Radecky:1981fn}%
  \BibitemOpen
  \bibfield  {author} {\bibinfo {author} {\bibfnamefont {G.}~\bibnamefont
  {Radecky}}, \bibinfo {author} {\bibfnamefont {V.}~\bibnamefont {Barnes}},
  \bibinfo {author} {\bibfnamefont {D.}~\bibnamefont {Carmony}}, \bibinfo
  {author} {\bibfnamefont {A.}~\bibnamefont {Garfinkel}}, \bibinfo {author}
  {\bibfnamefont {M.}~\bibnamefont {Derrick}},  \emph {et~al.},\ }\href
  {\doibase 10.1103/PhysRevD.25.1161, 10.1103/PhysRevD.26.3297} {\bibfield
  {journal} {\bibinfo  {journal} {Phys.Rev.}\ }\textbf {\bibinfo {volume}
  {D25}},\ \bibinfo {pages} {1161} (\bibinfo {year} {1982})}\BibitemShut
  {NoStop}%
\bibitem [{\citenamefont {Kitagaki}\ \emph {et~al.}(1990)\citenamefont
  {Kitagaki}, \citenamefont {Yuta}, \citenamefont {Tanaka}, \citenamefont
  {Yamaguchi}, \citenamefont {Abe} \emph {et~al.}}]{Kitagaki:1990vs}%
  \BibitemOpen
  \bibfield  {author} {\bibinfo {author} {\bibfnamefont {T.}~\bibnamefont
  {Kitagaki}}, \bibinfo {author} {\bibfnamefont {H.}~\bibnamefont {Yuta}},
  \bibinfo {author} {\bibfnamefont {S.}~\bibnamefont {Tanaka}}, \bibinfo
  {author} {\bibfnamefont {A.}~\bibnamefont {Yamaguchi}}, \bibinfo {author}
  {\bibfnamefont {K.}~\bibnamefont {Abe}},  \emph {et~al.},\ }\href {\doibase
  10.1103/PhysRevD.42.1331} {\bibfield  {journal} {\bibinfo  {journal}
  {Phys.Rev.}\ }\textbf {\bibinfo {volume} {D42}},\ \bibinfo {pages} {1331}
  (\bibinfo {year} {1990})}\BibitemShut {NoStop}%
\bibitem [{\citenamefont {Galster}\ \emph {et~al.}(1972)\citenamefont {Galster}
  \emph {et~al.}}]{Galster:1972rh}%
  \BibitemOpen
  \bibfield  {author} {\bibinfo {author} {\bibfnamefont {S.}~\bibnamefont
  {Galster}} \emph {et~al.},\ }\href@noop {} {\bibfield  {journal} {\bibinfo
  {journal} {Phys. Rev.}\ }\textbf {\bibinfo {volume} {D5}},\ \bibinfo {pages}
  {519} (\bibinfo {year} {1972})}\BibitemShut {NoStop}%
\bibitem [{\citenamefont {Lalakulich}\ \emph {et~al.}(2006)\citenamefont
  {Lalakulich}, \citenamefont {Paschos},\ and\ \citenamefont
  {Piranishvili}}]{Lalakulich:2006sw}%
  \BibitemOpen
  \bibfield  {author} {\bibinfo {author} {\bibfnamefont {O.}~\bibnamefont
  {Lalakulich}}, \bibinfo {author} {\bibfnamefont {E.~A.}\ \bibnamefont
  {Paschos}}, \ and\ \bibinfo {author} {\bibfnamefont {G.}~\bibnamefont
  {Piranishvili}},\ }\href {\doibase 10.1103/PhysRevD.74.014009} {\bibfield
  {journal} {\bibinfo  {journal} {Phys.Rev.}\ }\textbf {\bibinfo {volume}
  {D74}},\ \bibinfo {pages} {014009} (\bibinfo {year} {2006})},\ \Eprint
  {http://arxiv.org/abs/hep-ph/0602210} {arXiv:hep-ph/0602210 [hep-ph]}
  \BibitemShut {NoStop}%
\bibitem [{\citenamefont {Graczyk}\ \emph {et~al.}(2009)\citenamefont
  {Graczyk}, \citenamefont {Kielczewska}, \citenamefont {Przewlocki},\ and\
  \citenamefont {Sobczyk}}]{Graczyk:2009qm}%
  \BibitemOpen
  \bibfield  {author} {\bibinfo {author} {\bibfnamefont {K.}~\bibnamefont
  {Graczyk}}, \bibinfo {author} {\bibfnamefont {D.}~\bibnamefont
  {Kielczewska}}, \bibinfo {author} {\bibfnamefont {P.}~\bibnamefont
  {Przewlocki}}, \ and\ \bibinfo {author} {\bibfnamefont {J.}~\bibnamefont
  {Sobczyk}},\ }\href {\doibase 10.1103/PhysRevD.80.093001} {\bibfield
  {journal} {\bibinfo  {journal} {Phys.Rev.}\ }\textbf {\bibinfo {volume}
  {D80}},\ \bibinfo {pages} {093001} (\bibinfo {year} {2009})},\ \Eprint
  {http://arxiv.org/abs/0908.2175} {arXiv:0908.2175 [hep-ph]} \BibitemShut
  {NoStop}%
\bibitem [{\citenamefont {Adler}\ \emph {et~al.}(1974)\citenamefont {Adler},
  \citenamefont {Nussinov},\ and\ \citenamefont {Paschos}}]{Adler:1974qu}%
  \BibitemOpen
  \bibfield  {author} {\bibinfo {author} {\bibfnamefont {S.~L.}\ \bibnamefont
  {Adler}}, \bibinfo {author} {\bibfnamefont {S.}~\bibnamefont {Nussinov}}, \
  and\ \bibinfo {author} {\bibfnamefont {E.~A.}\ \bibnamefont {Paschos}},\
  }\href {\doibase 10.1103/PhysRevD.9.2125, 10.1103/PhysRevD.10.1669.3}
  {\bibfield  {journal} {\bibinfo  {journal} {Phys.Rev.}\ }\textbf {\bibinfo
  {volume} {D9}},\ \bibinfo {pages} {2125} (\bibinfo {year}
  {1974})}\BibitemShut {NoStop}%
\bibitem [{\citenamefont {Paschos}\ \emph {et~al.}(2007)\citenamefont
  {Paschos}, \citenamefont {Schienbein},\ and\ \citenamefont
  {Yu}}]{Paschos:2007pe}%
  \BibitemOpen
  \bibfield  {author} {\bibinfo {author} {\bibfnamefont {E.~A.}\ \bibnamefont
  {Paschos}}, \bibinfo {author} {\bibfnamefont {I.}~\bibnamefont {Schienbein}},
  \ and\ \bibinfo {author} {\bibfnamefont {J.-Y.}\ \bibnamefont {Yu}},\
  }\href@noop {} {\  (\bibinfo {year} {2007})},\ \Eprint
  {http://arxiv.org/abs/0704.1991} {arXiv:0704.1991 [hep-ph]} \BibitemShut
  {NoStop}%
\bibitem [{\citenamefont {Bolognese}(1978)}]{Bolognese:1978yz}%
  \BibitemOpen
  \bibfield  {author} {\bibinfo {author} {\bibfnamefont {T.}~\bibnamefont
  {Bolognese}},\ }\href@noop {} {\enquote {\bibinfo {title} {{Etude des
  interactions d'antineutrinos avec production d'un pion en courant
  charg\'e}},}\ } (\bibinfo {year} {1978}),\ \bibinfo {note} {{PhD thesis (in
  French). Universit\'e Louis Pasteur de Strasbourg. Report number CRN/HE
  78-22.}}\BibitemShut {Stop}%
\bibitem [{\citenamefont {Leitner}\ \emph {et~al.}(2009)\citenamefont
  {Leitner}, \citenamefont {Buss}, \citenamefont {Alvarez-Ruso},\ and\
  \citenamefont {Mosel}}]{Leitner:2008ue}%
  \BibitemOpen
  \bibfield  {author} {\bibinfo {author} {\bibfnamefont {T.}~\bibnamefont
  {Leitner}}, \bibinfo {author} {\bibfnamefont {O.}~\bibnamefont {Buss}},
  \bibinfo {author} {\bibfnamefont {L.}~\bibnamefont {Alvarez-Ruso}}, \ and\
  \bibinfo {author} {\bibfnamefont {U.}~\bibnamefont {Mosel}},\ }\href
  {\doibase 10.1103/PhysRevC.79.034601} {\bibfield  {journal} {\bibinfo
  {journal} {Phys.Rev.}\ }\textbf {\bibinfo {volume} {C79}},\ \bibinfo {pages}
  {034601} (\bibinfo {year} {2009})},\ \Eprint {http://arxiv.org/abs/0812.0587}
  {arXiv:0812.0587 [nucl-th]} \BibitemShut {NoStop}%
\bibitem [{\citenamefont {Aguilar-Arevalo}\ \emph {et~al.}(2009)\citenamefont
  {Aguilar-Arevalo} \emph {et~al.}}]{AguilarArevalo:2008yp}%
  \BibitemOpen
  \bibfield  {author} {\bibinfo {author} {\bibfnamefont {A.}~\bibnamefont
  {Aguilar-Arevalo}} \emph {et~al.} (\bibinfo {collaboration} {MiniBooNE
  Collaboration}),\ }\href {\doibase 10.1103/PhysRevD.79.072002} {\bibfield
  {journal} {\bibinfo  {journal} {Phys.Rev.}\ }\textbf {\bibinfo {volume}
  {D79}},\ \bibinfo {pages} {072002} (\bibinfo {year} {2009})},\ \Eprint
  {http://arxiv.org/abs/0806.1449} {arXiv:0806.1449 [hep-ex]} \BibitemShut
  {NoStop}%
\end{thebibliography}%

\end{document}